\documentclass[12pt]{article}
\usepackage[margin=1in]{geometry}  % set the margins to 1in on all sides
\usepackage[singlespacing]{setspace}
\usepackage{epsfig}
\usepackage{graphics,graphicx,color}
\usepackage{amssymb,amsfonts,amsthm,amsmath}
\usepackage{authblk}
\usepackage{multirow}
\usepackage{mathrsfs}
\usepackage{enumerate}
\usepackage[authoryear]{natbib}
\usepackage{float}
\usepackage{algorithm}
\usepackage{xcolor}
\usepackage{algpseudocode}
\usepackage{relsize}
\usepackage{subcaption} 
\usepackage{multirow}
\usepackage{url}
\usepackage[colorlinks=true, linkcolor=blue, citecolor=blue]{hyperref}
\usepackage{bm}
\usepackage{comment}
\usepackage{dsfont}
\usepackage{medmath}
\usepackage{subcaption}  % Required for subfigure environment

\usepackage[dvipsnames]{xcolor}

\newcommand{\bx}{\mbox{\boldmath{$x$}}}

\graphicspath{ {./Figures/} }

\title{Emulators for Large-scale Computer Experiments with Quantitative and Qualitative Inputs}

\author[1]{Anita Shahrokhian}
\author[2]{Youngdeok Hwang} 
\author[1]{C. Devon Lin}

\affil[1]{Department of Mathematics and Statistics, Queen's University}
\affil[2]{Paul H. Chook Department of Information Systems and Statistics, Baruch College, The City University of New York}
\begin{document}
	
	\date{}
	\maketitle
	
	% REQUIRED
	\begin{abstract}
		Computer experiments with both quantitative and qualitative inputs have become common across various areas. 
		However, constructing accurate and computationally efficient emulators for such experiments at large scales remains a significant challenge. 
		We propose a novel, scalable framework for emulating computer experiments with mixed inputs. 
		Our approach is based on a new covariance function integrating additive Gaussian processes 
		to handle the mixed inputs, with Vecchia approximation for scalability. 
		We demonstrate that methods for large-scale computer experiments can be effectively extended when paired with our proposed modeling framework using simulation studies and a real  structural engineering application. 
	\end{abstract}

%	
%	% REQUIRED
%	\begin{keywords}
%		Big data, Gaussian process,  Kriging,  Surrogate, Uncertainty quantification, Vecchia approximation.
%	\end{keywords}
%	
%	% REQUIRED
%	\begin{MSCcodes}
%		62-08, 62M30, 62P30
%	\end{MSCcodes}
%	
%	
	
	\section{Introduction}
	
	% \begin{enumerate}
		%     \item Explain why computer experiments are important. keep it not-too-long.
		%     \item Mixed inputs
		%     \item Large scale
		%     \item teaser for the new proposal, and why it is an important contribution
		% \end{enumerate}
	
	Computer experiments have become an essential tool to drive new discovery and innovation across a wide spectrum of scientific and engineering, for combustion \cite{zhou2025region}, system biology \cite{Hwang03042025}, energy storage \cite{escalante2021uncertainty}, environment \cite{holthuijzen2025synthesizing}, among many.
	These computer models allow researchers to overcome the inherent limitation of physical experiments, such as prohibitive costs, physical infeasibility, or ethical constraints \cite{gramacy2020surrogates}.

	Computer experiments increasingly involve mixed inputs, both quantitative and qualitative inputs, with wider applications of computer models of real-world systems \citep{shahrokhiana2025active}. It is also a common practice to convert quantitative variables to a set of fixed values and treat them as qualitative variables when it is difficult to change the experimental settings. For instance, \cite{qian2008gaussian}'s data center thermal flow case study had three qualitative factors, 
	\cite{deng2017additive} and \cite{xiao2021ezgp} considered Young's modulus of columns and other qualitative variables for embankments construction, and in the high-performance computing application studied by \cite{cai2024adaptive} and \cite{shahrokhian2025adaptive}, Input/Output (IO) operation mode  and IO scheduler are qualitative input variables.

	With its increased importance and broader use, there have been efforts to incorporate mixed inputs in
	statistical surrogate models. A metric space needs to be introduced for qualitative variables using
	an appropriate covariance structure, such as a multiplicative form \citep{qian2008gaussian, zhou2011simple}
	or an additive form \citep{deng2017additive, xiao2021ezgp}. 
	When the number of qualitative variables increases, however, 
	the computation becomes challenging because the number of covariance parameters increases rapidly.
	To keep the computation and estimation manageable, 
	rigid assumptions are often placed on the covariance structure, 
	which also adversarially affects the model performance.

	The use of statistical emulators for large-scale computer experiments poses significant computational challenges.
	As dataset sizes grow, conventional Gaussian process (GP)-based approaches become computationally prohibitive, with computational and memory requirements scaling as $\mathcal{O}(n^3)$ and $\mathcal{O}(n^2)$, respectively, for $n$ observations. 
	The Vecchia approximation \citep{vecchia1988estimation,katzfuss2020vecchia} enables scalable estimation with faster likelihood evaluation for parameter inference in large datasets. Meanwhile, the abundance of data in massive datasets has allowed response surfaces to be modeled with localized structures,  giving rise to methods such as the localized approximate  GP \citep[laGP,][]{gramacy2015local} and TwinGP \citep{vakayil2024global}.
	
	Despite recent advances, scalable GP modeling for massive datasets with mixed inputs remains under-explored. Existing mixed-input
	emulation methods, such as the easy-to-interpret GPs (EzGPs) of \cite{xiao2021ezgp} and latent variable GPs (LvGPs) of \cite{zhang2020latent}, offer flexible modeling frameworks for mixed inputs, but they face  computational challenges when applied to large-scale datasets.
	To scale LvGPs, \cite{yerramilli2023fully} proposed approximations for both the models and its fully Bayesian hyperparameter inference. They concluded that the Bayesian treatment offered limited benefits, though this was demonstrated only on small datasets (of size $\leq 100$). On the other hand, Vecchia-based approximations  and localized GP methods offer computationally affordable modeling, they have not yet
	been extended to handle mixed inputs. To bridge this gap, we introduce a novel covariance function that incorporates qualitative factors into a computationally scalable framework. Our approach, building on recent advances, addresses the key challenges of modeling large-scale datasets with mixed inputs by enabling efficient prediction and full uncertainty quantification.
	
	The remainder of this article is organized as follows. 
	Section \ref{sec:background} provides a review of EzGP and associated methods, 
	and the scaled Vecchia approximation. A new covariance function for mixed inputs is introduced in Section \ref{sec:covariance-function}. 
	Section \ref{sec:extending-lagp-twingp}
	demonstrates how the proposed model adapts 
	the two prominent methods for large-scale computer experiments.
	Section \ref{sec:numerical-illustration}
	presents several numerical studies 
	to demonstrate the proposed method. 
	Section \ref{sec:case-study} presents a case study with a real application  in structural engineering.
	Section \ref{sec:conclusion} concludes
	the paper.

	\section{Background} \label{sec:background}
	
	% review EzGP
	% scaled Vecchia approximation
	% 
	
	In this section, we review the methodologies that our approach is based upon. 
	We first review the GP models for mixed inputs proposed by \cite{xiao2021ezgp}, followed by the scaled Vecchia approximation introduced by \cite{katzfuss2022scaled}  for scalable analysis and emulation of large-scale computer experiments. 
	
	\subsection{Easy-to-Interpret GP models} \label{subsec:ezgp}
	% This section reviews the methods proposed by \cite{xiao2021ezgp} for modeling computer experiments with both quantitative and qualitative inputs, as well as the scaled Vecchia approximation introduced by \cite{katzfuss2022scaled} for scalable analysis and emulation of large-scale computer experiments. 
	% We first 
	% review the easy-to-interpret (EzGP), the efficient EzGP (EEzGP), and the localized EzGP (LEzGP) models. 
	
	Consider a computer experiment with $p$ quantitative variables $x^{(1)}, x^{(2)}, 
	\ldots, x^{(p)}$ and $q$ qualitative factors $z^{(1)}, z^{(2)}, 
	\ldots, z^{(q)} $ where $z^{(h)}$ has $m_h$ levels,  for $h=1,\ldots, q$. Given the $n$ responses $y_1,\ldots, y_n$ and $n$ 
	inputs $\bm{w}_1,\ldots, \bm{w}_n$ where $\bm{w}_i = (\bm{x}_i^\top,\bm{z}_i^\top)^\top$,  $\bm{x}_i = (x_{i1},\ldots, x_{ip})^\top$ and 
	$\bm{z}_i = (z_{i1},\ldots, z_{iq})^\top$, to model the relationship between output $Y$ and an input $\bm{w} = (\bm{x}^\top,\bm{z}^\top)^\top$,  \cite{xiao2021ezgp} proposed the model
	\begin{align}\label{ADGP}
		Y(\bm{x},\bm{z})=\mu+G_0(\bm{x})+G_{{z}^{(1)}}(\bm{x})+\ldots+G_{{z}^{(q)}}(\bm{x}),
	\end{align}
	where $\mu$ is a constant mean, $G_0$ and $G_{{z}^{(h)}} $'s are independent GPs with mean zero and covariance function $\bm{\phi}_0$ and $\bm{\phi}_h$, $h=1,\ldots,q$.  \cite{xiao2021ezgp} assumes that  $G_0$ is a standard GP with only quantitative inputs $\bm{x}$ in $\bm{\phi}_0$ given by, 
	\begin{align}\label{EzGpcov1}  \bm{\phi}_0(\bm{x}_i,\bm{x}_j|\bm{\theta}^{(0)})=\sigma_0^2\exp\{-\sum_{k=1}^{p}\theta_k^{(0)}(x_{ik}-x_{jk})^2\},
	\end{align}
	\noindent where all correlation parameters $\bm{\theta}^{(0)} =(\theta_1^{(0)},\ldots,\theta_p^{(0)})^\top$ are positive. For $h=1, \ldots,q$, the covariance function $\bm{\phi}_h$ can be defined as
	\begin{align*}
		\begin{split}
			\bm{\phi}_h((\bm{x}_i^\top,z_{ih})^\top,(\bm{x}_j^\top,z_{jh})^\top|\bm{\Theta}^{(h)})&=\sigma_h^2\exp\{-\sum_{k=1}^{p}\theta^{(h)}_{kz_{ih}z_{jh}}(x_{ik}-x_{jk})^2\}\tau^{(h)}_{z_{ih}z_{jh}},\\
		\end{split}
	\end{align*}
	\noindent where $\theta^{(h)}_{kz_{ih}z_{jh}}$'s are the correlation parameters for the pair of levels $z_{ih}$ and $z_{jh}$ in ${z}^{(h)}$, $\tau^{(h)}_{z_{ih}z_{jh}}$ is the correlation between the two levels and $\sigma_h^2$ is the variance for ${z}^{(h)}$. 
	This formulation leads to a large number of parameters and often suffers from computational challenges. 
	To avoid this over-parametrization and improve the interpretability, 
	\cite{xiao2021ezgp} proposed a more efficient form  of $G_{{z}^{(h)}}$   given by
	
	{\footnotesize
		\begin{align}\label{EzGpcov2}
			\begin{split}
				\bm{\phi}_h((\bm{x}_i^\top,z_{ih})^\top,(\bm{x}_j^\top,z_{jh})^\top|\bm{\Theta}^{(h)})&=\sum_{l_h=1}^{m_h}\sigma_h^2\exp\{-\sum_{k=1}^{p}\theta^{(h)}_{kl_h}(x_{ik}-x_{jk})^2\}\bm{\mathds{1}}(z_{ih}=z_{jh}\equiv l_h),\\
			\end{split}
		\end{align}
	}
	
	\noindent where % $z_{ih}$ and $z_{jh}$ are the $i$th and $j$th value of ${z}^{(h)}$, 
	%$m_h$ is the number of the levels of ${z}^{(h)}$, 
	$\bm{\Theta}^{(h)}=(\theta^{(h)}_{kl_h})_{p\times m_h}$ is the matrix for the correlation parameters, and the indicator function is 1 when $z_{ih}=z_{jh}\equiv l_h$ and 0 otherwise for $l_h=1,\ldots,m_h$. Therefore, from (\ref{EzGpcov1}) - (\ref{EzGpcov2}), the covariance function for the model in (\ref{ADGP}) is given by
	\begin{align}\label{EZGPcov}
		\begin{split}
			\bm{\phi}(\bm{w}_i,\bm{w}_j)&=\mbox{Cov}(Y(\bm{w}_i),Y(\bm{w}_j)) = \bm{\phi}_0(\bm{x}_i,\bm{x}_j|\bm{\theta}^{(0)})+\sum_{h=1}^{q}\bm{\phi}_h(\bm{x}_i,\bm{x}_j|\bm{\Theta}^{(h)}) \\
			& =\sigma_0^2\exp\{-\sum_{k=1}^{p}\theta_k^{(0)}(x_{ik}-x_{jk})^2\} 
			\\
			&\ \ \ \ +\sum_{h=1}^{q}\sum_{l_h=1}^{m_h}\bm{\mathds{1}}(z_{ih}=z_{jh}\equiv l_h)\sigma_h^2\exp\{-\sum_{k=1}^{p}\theta^{(h)}_{kl_h}(x_{ik}-x_{jk})^2\}.
		\end{split}
	\end{align}
	This covariance function has $2+p+q+p\sum_{h=1}^{q}m_h$ parameters $\mu$, $\sigma_0^2$, $\theta_k^{(0)}$, $\sigma_h^2$, and $\theta_{kl_h}^{(h)}$ for  $h=1,\ldots,q,~k=1,\ldots,p$ and $l_h=1,\ldots,m_h$, which can be estimated using the {maximum likelihood estimation} (MLE) method. 
	Under the GP model in (\ref{ADGP}), 
	the parameters can be found by minimizing the negative log-likelihood given by
	\begin{align}\label{loglikelihood}
		\log|\bm{\Phi}|+(\bm{y}-\mu \bm{\mathds{1}}_n)^{\top}\bm{\Phi}^{-1}(\bm{y}-\mu \bm{\mathds{1}}_n),
	\end{align} 
	with 
	$\bm{y}=(y_1,\ldots,y_n)^\top$, 
	$\bm{\mathds{1}}_n$ being a  vector of $n$ ones, 
	$\bm{\Phi}=(\bm{\phi}(\bm{w}_i,\bm{w}_j))_{n\times n}$ with the covariance given in \eqref{EZGPcov},
	after dropping the   terms that are not dependent on the parameters.
	We denote the parameters   in \eqref{EZGPcov} by
	$\bm{\sigma}^2=(\sigma_0^2,\ldots,\sigma^2_q)^\top$,  
	$\bm{\Theta}=(\bm{\theta}^{(0)},\bm{\Theta}^{(1)},\ldots,\bm{\Theta}^{(q)})$ where 
	$\bm{\theta}^{(0)}=(\theta_k^{(0)})_{ p \times 1}$, and $\bm{\Theta}^{(h)}=(\theta_{kl_h}^{(h)})_{p\times m_h}$.

	For a given $\bm{\sigma}^2$ and $\bm{\Theta}$, the maximum likelihood estimate of $\mu$ is 
	\begin
	{align}\label{mu}
	\hat{\mu}=(\bm{\mathds{1}}_n^\top\bm{\Phi}^{-1}\bm{\mathds{1}}_n)^{-1}\bm{\mathds{1}}_n^\top\bm{\Phi}^{-1}\bm{y},
\end{align}
and $\bm{\sigma}^2$ and $\bm{\Theta}$ can be obtained from
\begin{align*}
	\{\hat{\bm{\sigma}}^2,\hat{\bm{\Theta}}\}=\operatorname*{argmin}_{\bm{\sigma}^2,\bm{\Theta}}\{\log|\bm{\Phi}|+\bm{y}^\top\bm{\Phi}^{-1}\bm{y}-(\bm{\mathds{1}}_n^\top\bm{\Phi}^{-1}\bm{\mathds{1}}_n)^{-1}(\bm{\mathds{1}}_n^\top\bm{\Phi}^{-1}\bm{y})^2\},
\end{align*}
using an optimization algorithm, such as \texttt{rgenoud} in R \citep{r-genoud}. 
Let $Y^{*}=Y(\bm{w}^*)$ denote the predicted $Y$ at a new input $\bm{w}^*$. Given $\hat{\mu}$, $\hat{\bm{\sigma}}^2,$ and $\hat{\bm{\Theta}}$, the predictive mean and the predictive variance at $\bm{w}^{*}$ are  
\begin{align}\label{ARSDC_dist}
	\begin{split}
		\mbox{E}(Y^{*}|\bm{y})&=\mu_{*}=\hat{\mu}+\bm{r}_0^\top\bm{\Phi}^{-1}(\bm{y}-\hat{\mu}\bm{\mathds{1}}_n),\\
		\mbox{Var}(Y^{*}|\bm{y})&=\sigma^{{2}}_{*}=\sum_{i=0}^{q}\sigma^2_{i}-\bm{r}_0^\top\bm{\Phi}^{-1}\bm{r}_0+\frac{(1-\bm{\mathds{1}}_n^\top\bm{\Phi}^{-1}\bm{r}_0)^2}{\bm{\mathds{1}}_n^\top\bm{\Phi}^{-1}\bm{\mathds{1}}_n},
	\end{split}
\end{align}
respectively, where $\bm{r}_0$ is the covariance vector of $\bm{\Phi}(\bm{w}^{*},\bm{w}_i)_{n\times 1}$ for $i=1,\ldots,n$. The covariance function of EzGP in (\ref{EZGPcov}) considers the distinct correlation parameters for $\theta_{kl_h}^{(h)}$ to scale the quantitative factors separately. 
To further simplify the computation, \cite{xiao2021ezgp} proposed another covariance function called EEzGP, which uses a single correlation parameter $\theta_{l_h}^{(h)}$ to scale all quantitative factors equally, rather than considering separate parameters for each quantitative factor. The covariance function for EEzGP is given by 
\begin{align}\label{EEzGPcov}
	\begin{split}
		\phi(\bm{w}_i,\bm{w}_j)&= \mbox{Cov}(Y(\bm{w}_i),Y(\bm{w}_j))\\
		&=\sigma_0^2\exp\{-\sum_{k=1}^{p}\theta_k(x_{ik}-x_{jk})^2\}\\
		&+\sum_{h=1}^{q}\sum_{l_h=1}^{m_h}\bm{\mathds{1}}(z_{ih}=z_{jh}\equiv l_h)\sigma_h^2\exp\{-\sum_{k=1}^{p}\theta^{(h)}_{l_h}(x_{ik}-x_{jk})^2\}.
	\end{split}
\end{align}
This model has $2+p+q+\sum_{h=1}^{q}m_h$, which is much smaller than the number of parameters in the EzGP model.

However, both the EzGP and EEzGP models become computationally prohibitive for large $n$ as 
these models have computational complexity of $O(n^3)$ and 
memory space complexity of $O(n^2)$ respectively with the training dataset size $n$. 
% where $n$ denotes the size of the training dataset, 
To mitigate this issue, \cite{xiao2021ezgp} introduced the Localized EzGP (LEzGP) model, 
which works by selecting a sensible subset of training data for fitting the EEzGP or EzGP model 
conditional on a target input.

Consider an input vector $\bm{w} = (\bm{x}^\top, \bm{z}^\top)^\top$ and a target input $\bm{w}^{*} = (\bm{x}^{*\top}, \bm{z}^{*\top})^\top$. Let $N_z(\bm{w}, \bm{w}^{*})$ denote the number of matching levels between the qualitative inputs $\bm{z}$ and $\bm{z}^{*}$. For example, if $\bm{z} = (1, 2, 3)^\top$ and $\bm{z}^{*} = (3, 2, 1)^\top$, then $N_z(\bm{w}, \bm{w}^{*}) = 1$, since only the second level matches. The LEzGP model defines a tuning parameter $n_s$ to be 
the minimum number of matching qualitative levels required for subset selection. Only inputs satisfying $N_z(\bm{w}, \bm{w}^{*}) \geq n_s$ are selected. For a given $\bm{w}^{*}$, this condition defines a subset of inputs with at least $n_s$ matching qualitative levels. The choice of $n_s$ is critical, as it directly influences  the selection and the subset size. 
The following steps summarize the LEzGP model given the value of $n_s$:
\begin{enumerate}
	%    \item Select a tuning parameter $n_s$, which determines the minimum number of matching qualitative levels required for subset selection.
	\item For a target input $\bm{w}^*$, identify all inputs $\bm{w}_i$ (where $i = 1, \dots, n$) that satisfy the matching criterion:
	$N_z(\bm{w}_i, \bm{w}^*) \geq n_s.$
	Denote this selected subset as $\bm{K}_s$.
	
	\item Using the subset $\bm{K}_s$, apply either:
	the EzGP model in (\ref{EZGPcov}) or  the EEzGP model in  (\ref{EEzGPcov}),
	to predict the response at the target input $\bm{w}^*$ from Step 1.
\end{enumerate}
For instance, consider an experiment with $p=1$ and $q=3$ each having 3 levels, where the design matrix is 
\begin{align*}
	\bm{w}&=\begin{bmatrix}
		0.5 & 1&1&1\\0.6& 3&2&3\\0.7& 2&1&3\\0.8& 2&2&3\\\end{bmatrix}, \end{align*}
and the target input $\bm{w}^*=(0.3,1,2,3)^\top$.  Let $n_s=2$.  The subset $\bm{K}_s$ is selected such that at least $n_s$ matching qualitative levels between each row of $\bm{K}_s$ and the target input $\bm{w}^*$. 
Hence, both $(0.6, 3,2,3)^{\top}$ and $(0.8, 2,2,3)^{\top}$ will be selected for $\bm{K}_s$.

\subsection{The Vecchia approximation} \label{subsec:vecchia-approximation}

% review scaled vecchia approximation 

The Vecchia approximation is a computationally efficient method for approximating large covariance matrices in spatial statistics and GP modeling \citep{vecchia1988estimation}. By leveraging conditional independence in GP, it decomposes the joint likelihood into a product of low-dimensional conditional distributions, significantly reducing computational complexity from $\mathcal{O}(n^3)$ to approximately $\mathcal{O}(n m^2)$, where $m \ll n$ is the size of a carefully selected conditioning set. This approach has been extended to  fast emulation for large-scale computer experiments by \cite{katzfuss2022scaled}. They employed the Vecchia approximation to construct efficient emulators by transforming the input space, scaling each input variable according to its influence on the computer model's responses.   

Consider an $n$-run computer experiment with $p$ quantitative variables. The data includes inputs  $\bm{x}_i = (x_{i1},\ldots, x_{ip})^\top$ and the corresponding response $y_i=y(\bm{x}_i)$ for $i=1,\ldots, n$. Let  $\bm{y}=(y_1,\ldots,y_n)^\top$. Assume that $\bm{y}$ follows a GP with mean ${\mu}$, and covariance function $\bm{K}=({k}(\bm{x}_i,\bm{x}_j))_{i,j=1,\ldots,n}$. The Vecchia approximation approximates the joint density $p(\bm{y})=\prod_{i=1}^{n}p(y_i|y_1,\ldots,y_{i-1})$ by a product of univariate conditional densities, that is,  
\begin{align}\label{vecchiapp}
	\hat{p}(\bm{y})=\prod_{i=1}^{n}p(y_i|\bm{y}_{c(i)}),
\end{align}
where $c(i)\subset\{1,\ldots,i-1\}$ is a conditioning index of size $|c(i)|=\min(m_s,i-1)$ for $i=\{2,\ldots,n\}$ and $m_s$ is the size of the conditioning set. \cite{katzfuss2022scaled} commented that a relatively small size of $m_s\ll n$ in  (\ref{vecchiapp}) is sufficient for accurate approximation. Moreover, $p(y_i|\bm{y}_{c(i)})$ in (\ref{vecchiapp}) and all Gaussian distributions can be computed parallelly.  The accuracy of the Vecchia approximation is influenced by the choice of ordering $y_1, \ldots, y_n$ and the selection of the conditioning sets $c(i)$'s. \cite{katzfuss2022scaled} demonstrated that high accuracy can be achieved using the maximum-minimum distance (maximin) ordering and the nearest-neighbor (NN) conditioning.

\cite{katzfuss2022scaled} introduced a scaled Vecchia approximation  for fast emulation of computer models. They assume an anisotropic covariance function with a separate range parameter $\xi_l$ for the $l$th input dimension   known as automatic relevance determination,
$K(\bm{x}_i,\bm{x}_j) = \tilde{K}(q(\bm{x}_i,\bm{x}_j)) $
\noindent where $$ q(\bm{x}_i,\bm{x}_j) = \left (\sum_{l=1}^p  \left(\frac{x_{il}-x_{jl}}{\xi_l}\right)^2\right )^{1/2},$$ 
\noindent and $\tilde{K}$  can be any covariance function that is valid (i.e., strictly positive definite) in $R^p$.
The scaled Vecchia approximation applied the Vecchia approximation in a transformed input space, with each input scaled according to how strongly it relates to the computer-model
response. That is, they use the scaled inputs   
\begin{equation}\label{eq:scaled}
	\tilde{\bm{x}}=\left(\frac{x^{(1)}}{\xi_1},\ldots,\frac{x^{(p)}}{\xi_p} \right),
\end{equation}
\noindent where $\xi_i$'s are range parameters and $\frac{1}{\xi_l}$ is the relevance of  the $l$th input dimension. By scaling the inputs using the estimated range parameters, this approach controls the effect of input variables on the model. 
The scaled Vecchia approximation considers a maximin ordering and the NN conditioning of the scaled $\tilde{\bm{x}}$, not the original ${\bm{x}}$.  
%The parameters in the scaled Vecchia approximate include the parameter $\mu$ in the mean function and the parameters in the covariance function. 
Denote the mean and parameters including range parameters in the covariance in the scaled Vecchia approximation
by $\mu$ and $\bm{\theta}$. To estimate the parameters, \cite{katzfuss2022scaled} used the Fisher scoring algorithm in \cite{guinness2021gaussian} by maximizing the logarithm of the Vecchia likelihood in (\ref{vecchiapp}). 
Consider the approximate log joint density 
$l(\mu,\bm{\theta})=\log(\hat{p}_{\mu,\bm{\theta}})$, where $\hat{p}_{\mu,\bm{\theta}}$ is the Vecchia approximation in (\ref{vecchiapp}). 
Since computing the derivatives of the conditional densities in (\ref{vecchiapp}) can be computationally challenging,  replacing them by the joint distribution leads to
\begin{align}\label{Vecchialike}
	l({\mu},\bm{\theta})=\sum_{i=1}^{n}(\log{p_{\mu,\bm{\theta}}}(y_i,\bm{y}_{c(i)})-\log p_{\mu,\bm{\theta}}(\bm{y}_{c(i)})).
\end{align}
This formulation leverages the gradient and Fisher information for the Gaussian distribution, both of which can be computed using standard formulas, starting from an initial value of $\bm{\theta}^{(0)}$ such that, 
\begin{align}\label{fisherscoring}
	\bm{\theta}^{(k+1)}= \bm{\theta}^{(k)}+(\bm{M}^{(k)})^{-1}\bm{g}^{(k)},
\end{align}
where $\mathsmaller{\bm{g}^{(k)}=\frac{\partial l({\mu},\bm{\theta})}{\partial \bm{\theta}}|_{\bm{\theta}=\bm{\theta}^{(k)}}}$ and $\mathsmaller{\bm{M}^{(k)}=-E\{\frac{\partial^2 l({\mu},\bm{\theta})}{\partial\bm{\theta}\partial\bm{\theta^{\top}}}|_{\bm{\theta}=\bm{\theta}^{(k)}}\}}$
are computed based on (\ref{Vecchialike}). The details of the estimation can be found in \cite{guinness2021gaussian}. In \cite{katzfuss2022scaled}, the Fisher scoring iterations change with $\tilde{\bm{x}}$ in (\ref{eq:scaled}). To speed up the process, the parameter estimates are only updated at 
every doubling  step, i.e., $k=2,4,8,\dots$.

Given the estimated parameters, the Vecchia approximation provides predictions at  unobserved inputs $\bm{x}^{*}_1,\ldots,\bm{x}^{*}_{n_{*}}$ by applying the  Vecchia approximation to the joint density of $\bm{y}^{\text{\footnotesize all}}=(\bm{y},\bm{y}^{*})$, where $\bm{y}^*=(y_1^*,\ldots,y_{n_*}^*)^\top$ with ${y}_i^*=y(\bm{x}^{*}_i),~i=1,\ldots,n_*$.  This approximation is used to accurately compute the posterior distribution of $\bm{y}^*=(y_1^*,\ldots,y_{n_*}^*)^\top$ for large values of $n_*$.  The posterior predictive distribution is 
\begin{align}
	\hat{p}(\bm{y}^*|\bm{y})=\prod_{i=1}^{n_*}p(y^*_i|\bm{y}^{\text{\footnotesize all}}_{g^*(i)}),
\end{align}
where $g^{*}(i)$ includes $m_*$ points closes to $y^{*}_i$ in terms of scaled distance among the points are previously ordered in $\bm{y}^{\text{\footnotesize all}}$.

\section{A New Covariance Function} \label{sec:covariance-function}

In this section, we propose extending  the scaled Vecchia approximation  of \cite{katzfuss2022scaled} to computer experiments with quantitative and qualitative inputs.
The scaling method in \cite{katzfuss2022scaled} is carried out by multiplying each quantitative input variable $\bm{x}_l$ by its relevance parameter $1/{\xi}_l$, which is estimated from the anisotropic  covariance function using Fisher scoring. A naive approach of extending the scaled Vecchia approximation  to computer experiments with mixed inputs would be applying the scaling solely based on the covariance parameters associated with the quantitative inputs. In the case of EzGP or EEzGP, for example,  one can scale the $l$th quantitative variable by $\mathsmaller{\sqrt{\theta_l^{(0)}}}$ in (\ref{EzGpcov1}). 
This approach, however, only accounts for relevance parameters in the model component that uses purely quantitative inputs, it inherently overlooks the influence of qualitative factors.

To address this issue and incorporate the effects of both quantitative and qualitative inputs, we propose the following covariance function for (\ref{ADGP}):
%the EzGP covariance in (\ref{EZGPcov}) as follows,
\begin{align}\label{SEzGPcov}
	\begin{split}
		{K}(\bm{w}_i,\bm{w}_j)&=\textrm{Cov}(Y(\bm{w}_i),Y(\bm{w}_j))\\
		&=\sigma_0^2\exp\{-\sum_{k=1}^{p}\theta_k^{(0)}(x_{ik}-x_{jk})^2\}\\
		&+\sum_{h=1}^{q}\sum_{l_h=1}^{m_h}\mathds{1}(z_{ih}=z_{jh}\equiv l_h)\sigma_h^2\exp\{-\sum_{k=1}^{p}\theta_{k}(x_{ik}-x_{jk})^2\}.
	\end{split}
\end{align}
In contrast to the covariance function in (\ref{EZGPcov}), 
we set $\theta^{(h)}_{kz_{ih}z_{jh}}$ in (\ref{EZGPcov}) to be $\theta_{k}$. Our approach employs a GP that integrates both qualitative and quantitative inputs, but with correlation parameters that are invariant with respect to the qualitative variables. This constitutes a fundamental difference from the EEzGP covariance function in (\ref{EEzGPcov}), which instead assumes that parameters are independent of those of quantitative variables. Consequently, the assumptions underlying the EEzGP model also render it incompatible with an extension of the scaled Vecchia approximation to handle mixed inputs.

The SEzGP model involves $2 + 2p + q$ parameters, $\mu$, $\sigma_0^2$, $\theta_k^{(0)}$, $\sigma_h^2$, and $\theta_{k}$, where $h = 1, \ldots, q$ and $k = 1, \ldots, p$.  These parameters can be estimated using the MLE method as described in (\ref{loglikelihood}). The scaled Vecchia approximation is applied by using the scaled quantitative inputs $\tilde{\bx} = (\sqrt{\theta_1}x^{(1)},\sqrt{\theta_2}x^{(2)}, \ldots, \sqrt{\theta_p}x^{(p)})^{\top}$ assuming that the standardized input space $[0,1]^p$ for quantitative variables.

The analytical gradient expression to apply Fisher scoring  in (\ref{fisherscoring}) is
\begin{align*}
	-2\frac{\partial l(\bm{\sigma}^2,\bm{\Theta})}{\partial \bullet}=\mbox{tr}(\bm{K}^{-1}\frac{\partial \bm{K}}{\partial \bullet})-(\bm{y}-\hat{{\mu}})^\top\bm{K}^{-1}\frac{\partial \bm{K}}{\partial \bullet}\bm{K}^{-1}(\bm{y}-\hat{{\mu}}),
\end{align*}
where $\bm{\sigma}^2 = (\sigma_0^2,\sigma_1^2, \ldots, \sigma_q^2)$, $\bm{\Theta} = ( \theta_{1}^{(0)}, \ldots,\theta_{p}^{(0)},\theta_1, \ldots, \theta_p)$, with given $\hat{\mu}$ in (\ref{mu}).
Thus, for the SEzGP covariance function in (\ref{SEzGPcov}), and any $i,j=1,\ldots,n$, 
\begin{align*}
	\begin{split}
		\frac{\partial \bm{K}}{\partial \sigma_0^2}&=(\exp\{-\sum_{k=1}^{p}\theta^{(0)}_{k}(x_{ik}-x_{jk})^2\})_{n\times n},\\
		\frac{\partial \bm{K}}{\partial \sigma_h^2}&=(\sum_{l_h=1}^{m_h}\mathds{1}(z_{ih}=z_{jh}\equiv l_h)\exp\{-\sum_{k=1}^{p}\theta_{k}(x_{ik}-x_{jk})^2\})_{n\times n},\\
		\frac{\partial \bm{K}}{\partial \theta_{l}^{(0)}}&=(-\sigma_0^2(x_{il}-x_{jl})^2\exp\{-\sum_{k=1}^{p}\theta_{k}^{(0)}(x_{ik}-x_{jk})^2\})_{n\times n},\\
		\frac{\partial \bm{K}}{\partial \theta_{l} }&=(-\sum_{h=1}^{q}\sum_{l_h=1}^{m_h}\mathds{1}(z_{ih}=z_{jh}\equiv l_h)\sigma_h^2(x_{il}-x_{jl})^2\exp\{-\sum_{k=1}^{p}\theta_{k}(x_{ik}-x_{jk})^2\})_{n\times n}.\\
	\end{split}
\end{align*}
Using the scaled quantitative inputs $\tilde{\bx}$, the Vecchia likelihood in (\ref{Vecchialike}) can then be used to compute the gradient and the Fisher information matrix using the derivatives derived for the SEzGP covariance function. For the remainder of this paper, we call this method, which employs the scaled Vecchia approximation with the SEzGP covariance matrix, as SVA.

\section{Extending laGP and TwinGP} \label{sec:extending-lagp-twingp}

This section introduces extensions of two prominent methods for large-scale computer experiments to handle mixed inputs. The first method is the localized approximate GP (laGP) model proposed by \cite{gramacy2015local}, and the second is the twinGP model introduced by \cite{vakayil2024global}. These extensions are included as benchmark methods for comparison with the proposed approach.

\subsection{Extending laGP}
The  laGP models \citep{gramacy2015local} for fast emulation of large computer experiments use a family of sequential design schemes that dynamically define the support of a GP predictor based on a local subset of the data, and \cite{gramacy2015local} further derived expressions for fast sequential updating of all needed quantities as the local designs are built up iteratively. More specifically, let $\bm{X}_n$ be the $n$ inputs in the training data, and let $\bm{x}^*$ be the unobserved input for which the prediction will be made. Let $\bm{X}_m(\bm{x}^*) \subset \bm{X}_n$ be a set of $m$ points near $\bm{x}^*$, where $m$  is significantly smaller than $n$. For notational simplicity, we omit $\bm{x}^*$ and denote the subset   as $\bm{X}_m$.   The selection process of the local design with $m$ data points begins with an initial design of size $n_0$. A convenient choice of the initial design is the nearest neighbor  points close to   $\bm{x}^*$.  The follow-up points are selected via two  design criteria, the mean squared prediction error (MSPE) and active learning Cohn (ALC), aimed at enhancing prediction accuracy for each target test location by considering one step ahead and utilizing the currently available information \citep{gramacy2015local}. For $j = n_0, \ldots, m-1$, let $\bm{D}_j$ be the data including the inputs and outputs at step $j$.  The next input $\bm{x}_{j+1}$ is sequentially selected and the design is updated as $\bm{D}_{j+1} = \bm{D}_j \cup (\bm{x}_{j+1}, y_{j+1})$.  Given $\bm{x}^*$ and the current design $\bm{D}_j$, the MSPE criterion selects the next input $\bm{x}_{j+1}$ that maximizes
\begin{align}
	\begin{split}
		\text{MSPE}(\bm{x}_{j+1},\bm{x}^*)&=E\{[\bm{Y}(\bm{x}^*)-\mu_{j+1}(\bm{x}^*;\hat{\bm{\theta}}_{j+1})]^2|\bm{D}_j(\bm{x}^*)\}\\ 
		&\approx {V}_j(\bm{x}^*|\bm{x}_{j+1};\hat{\bm{\theta}}_j)+(\frac{\partial\mu_{j}(\bm{x}^*;\bm{\theta})}{\partial\bm{\theta}}|_{\bm{\theta}=\hat{\bm{\theta}}_j})^2/{G}_{j+1}(\hat{\bm{\theta}}_j),  \label{MSPE}
	\end{split}
\end{align}
\noindent where 
\[
V_j(\bm{x}^* \mid \bm{x}_{j+1}; \hat{\bm{\theta}}_j)
=
\frac{\psi_j}{j-2}
\, v_{j+1}(\bm{x}^*; \hat{\bm{\theta}}_j),
\]
with
\[
\psi_j
=
\bm{Y}_j^\top \bm{K}_j^{-1} \bm{Y}_j,
\qquad
\bm{Y}_j
=
(y_1,\ldots,y_j)^\top,
\]
and
\[
v_{j+1}(\bm{x}^*; \hat{\bm{\theta}}_j)
=
\bm{K}_{j+1}(\bm{x}^*,\bm{x}^*)
-
\bm{k}_{j+1}^\top(\bm{x}^*)
\bm{K}_{j+1}^{-1}
\bm{k}_{j+1}(\bm{x}^*).
\]
Furthermore,
$\frac{\partial \mu_j(\bm{x}^*;\bm{\theta})}
{\partial \bm{\theta}}$
denotes the derivative of the predictive mean at $\bm{x}^*$, given $\bm{D}_j$, with respect to $\bm{\theta}$, and
$
G_{j+1}(\hat{\bm{\theta}}_j)
$
denotes the expected Fisher information matrix, which consists of the information from $\bm{D}_j$ together with the expected contribution from the future observation $y_{j+1}$ at $\bm{x}_{j+1}$. The analytic expressions for
$\mathsmaller{
	\left.
	\frac{\partial \mu_j(\bm{x}^*;\bm{\theta})}
	{\partial \bm{\theta}}
	\right|_{\bm{\theta}=\hat{\bm{\theta}}_j}}
$
and
$\mathsmaller{
	G_{j+1}(\hat{\bm{\theta}}_j)}
$
are provided in \cite{gramacy2015local} and are omitted here for brevity.

The approximation proposed  in (\ref{MSPE}) consists of two terms. The first term represents the updated predictive variance after augmenting $\bm{D}_j$ with the candidate point $\bm{x}_{j+1}$, conditional on $\hat{\bm{\theta}}_j$. The second term accounts for the uncertainty arising from parameter estimation and involves two components:
$\mathsmaller{
	\frac{\partial \mu_j(\bm{x}^*;\bm{\theta})}
	{\partial \bm{\theta}}}$ 
which is the derivative of the predictive mean with respect to $\bm{\theta}$, and
$\mathsmaller{
	G_{j+1}(\hat{\bm{\theta}}_j),
}$ which is the expected Fisher information matrix incorporating both the information from $\bm{D}_j$ and the anticipated contribution from the future observation $y_{j+1}$ at $\bm{x}_{j+1}$.

\cite{gramacy2015local} discussed that when the Fisher information is large, the second term in (\ref{MSPE}) can be ignored and the criterion reduces to $\mathsmaller{{V}_j(\bm{x}^*|\bm{x}_{j+1};\hat{\bm{\theta}}_j)}$ that is the new variance at $\bm{x}^*$ when $\bm{x}_{j+1}$ is added to the design. In other words, the goal is to maximize the  ALC criterion, and that can be updated from $j$ to $j+1$ quickly via partitioned inverse equations by \cite{barnett1979matrix} such that, 
{\small \begin{align}\label{ALC_org}
		\begin{split}
			\text{ALC}(\bm{\bm{x}^*},\bm{x}_{j+1})&=\bm{k}_{j+1}^\top(\bm{x}^*) \boldsymbol{g}_j(\bm{x}_j,\bm{x}_{j+1})
			\boldsymbol{g}_j(\bm{x}_j,\bm{x}_{j+1})^\top {v}_j(\bm{x}_{j+1})\bm{k}_{j+1}^\top(\bm{x}^*) \\+
			&2\bm{k}_{j+1}^\top(\bm{x}^*) \boldsymbol{g}_{j}
			(\bm{x}_{j},\bm{x}_{j+1})\boldsymbol{K}(\bm{x}_{j+1},\bm{x}^*)
			+\boldsymbol{K}(\bm{x}_{j+1},\bm{x}^*)^2/ {v}_{j}(\bm{x}_{j+1}),
		\end{split}
	\end{align}
}

\noindent where $$\boldsymbol{g}_j(\bm{x}_{j},\bm{x}_{j+1})={\boldsymbol{K}_j}^{-1} \bm{k}_{j}(\bm{x}_{j+1}) /{v}_{j}(\bm{x}_{j+1})$$
\noindent and $$ {v}_{j}(\bm{x}_{j+1})=\boldsymbol{K}(\bm{x}_{j+1},\bm{x}_{j+1})-
\bm{k}_{j}(\bm{x}_{j+1})^\top{\boldsymbol{K}_j}^{-1} \bm{k}_{j}(\bm{x}_{j+1}).$$
The mathematical details for both the MSPE and ALC criteria are thoroughly presented in \cite{gramacy2015local}.  The fast update of the ALC criterion in (\ref{ALC_org}) is computationally efficient when the MLE parameters are updated only once at the design's final stage. 
These methods have been implemented in their R package \texttt{laGP} \citep{gramacy2016lagp}, which includes efficient routines for both the MSPE criterion (\ref{MSPE}) and the ALC criterion (\ref{ALC_org}).

We extend the use of the ALC criterion in (\ref{ALC_org}) to large-scale computer experiments with mixed inputs by incorporating the SEzGP model in (\ref{SEzGPcov}). Specifically, our goal is to  sequentially select \( \bm{w}_{j+1} \) that maximizes (\ref{ALC_org}) using the covariance function in (\ref{SEzGPcov}). 
We call this method as {\em La} for the remainder of this article.

The \texttt{laGP} package only supports isotropic covariance functions and is therefore not directly applicable to the SEzGP model. As a practical contribution, we implement the ALC criterion for SEzGP to extend the \texttt{laGP} framework to mixed-input settings, although the implementation is computationally slower because of the more complex covariance structure. 
% \textcolor{MidnightBlue}{
	% We can emphasize our formulation makes things simpler, not just for the package. 
	% This part needs a change. It's not just package that necessitates the SEzGP. 
	% laGP needs a notion of ``neighbor'' to select the design subset. 
	% Vecchia approximation, laGP and twinGP rely on the neighborhood distance, so we need 
	% an integrated distance metric. Distance in qualitative factor is difficult to work with. We need to emphasize this point.}

Similar to laGP, the predictive locations can be handled in parallel. 
For each test location $\bm{w}^*$,  we start with $n_0$ initial design chosen by NN, and fix the MLE parameters, 
then employ the ALC criterion in (\ref{ALC_org}) to select the next points adaptively.
At the end of the adaptive design process, the MLE parameters are updated once, as recommended in laGP.

\subsection{Extending TwinGP}

The TwinGP approach proposed by \cite{vakayil2024global} enhances prediction in large-scale computer experiments by strategically selecting global  and local points. The selection of global points 
remains independent of the specific locations to be predicted. 

The TwinGP model's correlation function combines two components:  $\bm{G}$  for $g$ global points and $\bm{L}$ for $l$ local points, and it is, 
\begin{align}\label{twinGP3}
	\bm{R}(\bm{x}_a,\bm{x}_b)=(1-\lambda)\bm{G}(\bm{x}_a,\bm{x}_b)+\lambda \bm{L}(\bm{x}_a,\bm{x}_b),~\lambda\in[0,1],
\end{align}
where  $\lambda$ controls the proportion of the global and local points to build the correlation matrix $\bm{R}$. Let $m = g+l$, $\bm{X}_n$ be the inputs in the training data, and $\bm{X}_m \subset \bm{X}_n$ be the chosen $m$  points from the training data, then we have $\bm{G}_{mm}=\bm{G}(\bm{X}_m,\bm{X}_m)$ and $\bm{L}_{mm}=\bm{L}(\bm{X}_m,\bm{X}_m)$, and  the correlation matrix for the $m$ chosen points is 
\begin{align*}
	\bm{R}_{mm}=(1-\lambda)\bm{G}_{mm}+\lambda \bm{L}_{mm},~\lambda\in[0,1].
\end{align*}
For each predictive location $\bm{x}^*$, the $g$ global points $\bm{X}_g$ are  chosen from $\bm{X}_n$   via the Twinning approach,
and the $l$ local points $\bm{X}_l$ are selected via the nearest-neighbor method from the remaining points $\bm{X}_n\setminus \bm{X}_g$.  The selection of $\bm{X}_l$ depends on $\bm{x}^*$, whereas $\bm{X}_g$ is independent of the predictive location. The combined set of points for modeling is $\bm{X}_m(\bm{x}^*) = \bm{X}_g \cup \bm{X}_l$, hereafter denoted as $\bm{X}_m$ for brevity.
Let $\bm{Y}_m$ be the response vector corresponding to $\bm{X}_m$, partitioned as $\bm{Y}_m = (\bm{Y}_g, \bm{Y}_l)$ for $\bm{X}_g$ and $\bm{X}_l$, respectively. In addition, 
% $\bm{\mathds{1}}_m$ denotes a column vector of $m$ ones, and 
$\bm{I}_m$, $\bm{I}_g$, and $\bm{I}_l$ represent identity matrices of dimensions $m \times m$, $g \times g$, and $l \times l$, respectively. %\textcolor{MidnightBlue}{We need to mention that $g$ and $l$ are number of global and local points in a better way. I added them 6 lines above, but it's a bit messy in its current form. We said $m$ is the training data size. }

The predictive mean and predictive variance involve 
\begin{align}
	\begin{split}
		\bm{R}_{mm}+\eta \bm{I}_m&=\begin{bmatrix}
			\bm{R}_{gg}+\eta \bm{I}_g & \bm{R}_{gl}\\\bm{R}_{lg}& \bm{R}_{ll}+\eta \bm{I}_l
		\end{bmatrix},\\
		&=\begin{bmatrix}
			(1-\lambda) \bm{G}_{gg}+\lambda \bm{L}_{gg}+\eta \bm{I}_g & (1-\lambda) \bm{G}_{gl}+\lambda \bm{L}_{gl}\\
			(1-\lambda) \bm{G}_{lg}+\lambda \bm{L}_{lg}& (1-\lambda) \bm{G}_{ll}+\lambda \bm{L}_{ll}+\eta \bm{I}_l
		\end{bmatrix},
	\end{split}
\end{align}
where $\bm{R}_{gg}$ is independent of the predictive location $\bm{x}^*$, implying that $[\bm{R}_{gg}+\eta \bm{I}_g]^{-1}$  can be predetermined beforehand. For a given $\bm{x}^*$, we have
\begin{align}\label{twingp4}
	\begin{split}
		[\bm{R}_{mm}+\eta \bm{I}_m]^{-1}&=\begin{bmatrix}
			\bm{\Sigma}_{gg}^{-1}+\bm{\Sigma}_{gg}^{-1} \bm{R}_{gl}\bm{S}^{-1} \bm{R}_{lg}\bm{\Sigma}_{gg}^{-1} & -\bm{\Sigma}_{gg}^{-1} \bm{R}_{gl}\bm{S}^{-1}\\
			-\bm{S}^{-1} \bm{R}_{lg}\bm{\Sigma}_{gg}^{-1}& \bm{S}^{-1}
		\end{bmatrix},
	\end{split}
\end{align}
where $\bm{\Sigma}_{gg}=\bm{R}_{gg}+\eta\bm{I}_g$ and $\bm{S}=\bm{R}_{ll}-\bm{R}_{lg}\bm{\Sigma}_{gg}^{-1}\bm{R}_{gl}$. 
\cite{vakayil2024global} provided an approximation to estimate the parameters in $\bm{R}_{ll}$ when the input variables are quantitative. 
This approximation, however,  does not apply to the covariance function of EzGP, EEzGP, or SEzGP. Therefore, we introduce an extension of TwinGP for large-scale computer experiments with mixed inputs.
%, inspired by the ideas presented by \cite{vakayil2024global}. 

Our goal here is to select a subset of $m$ training points $\bm{W}_m $ from the training data $\bm{W}_n$. Similar to the TwinGP method, we consider $\bm{W}_g$ as the set of global points selected using the Twinning method proposed by \cite{vakayil2022data}. Then, for a given predictive location $\bm{w}^*$, the local points \( \bm{W}_l \) are identified using the nearest neighbor  method. Before applying both the TwinGP method and the nearest neighbor method, we adopt one-hot encoding   to convert the qualitative variables into numeric formats. One-hot coding represents a categorical variable with $k$ levels by $k$ binary indicator variables, where exactly one variable takes the value 1 and the others take the value 0. For example, consider \( p=1, q=2, \mbox{and}\) with $m_1=m_2=2$. Here we have 4 different level combinations, and hence by applying one-hot encoding for two-level qualitative variables, we have
\begin{align*}
	\begin{split}
		(\bm{w}_1^\top,\bm{w}_2^\top,\bm{w}_3^\top,\bm{w}_4^\top)^\top &=\begin{bmatrix}
			0.5 & 1&1\\0.6& 1&2\\0.7& 2&1\\0.8& 2&2\\\end{bmatrix}
		\Rightarrow \begin{bmatrix}0.5 & 1&0&1&0\\0.6& 1&0&0&1\\0.7& 0&1&1&0\\0.8& 0&1&0&1\\\end{bmatrix}.
	\end{split}
\end{align*}
% and for three-level qualitative factors, 
% \begin{align*}
	% \begin{split}
		% \bm{w}_i&=\begin{bmatrix}
			% 0.5 & 1&1&1\\0.6& 1&2&3\\0.7& 2&1&3\\0.8& 2&2&1\\\end{bmatrix}
		% \Rightarrow\begin{bmatrix}0.5 & 1&0&0&1&0&0&1&0&0\\0.6& 1&0&0&0&1&0&0&0&1\\0.7& 0&1&0&1&0&0&0&0&1\\0.8& 0&1&0&0&1&0&0&0&1\\\end{bmatrix}.
		% \end{split}
	% \end{align*}
\noindent This conversion is solely used for selecting the global points and the local points;
the original form of the inputs is retained for all other parts of the modeling.
% \textcolor{MidnightBlue}{(minimal or delete) I think that we can make a direct connection of this conversion and 
	% $\sum_{l_h=1}^{m_h}\mathds{1}(z_{ih}=z_{jh}\equiv l_h)$ from the earlier part. 
	% If $z_{ih}=z_{jh}\equiv l_h$, then $\mathds{1}(z_{ih}=z_{jh}\equiv l_h)$ becomes 1, and the distance in $z_{ih} = z_{jh}$ is 0. So when there is a mismatch in categorical factors, we add an attenuation.
	% Devon: but this conversion does not use for modeling so it is not connected to the covariance matrix}
% , the original form of the inputs is used.  
% We have $\bm{W}_m = \bm{W}_g \cup \bm{W}_l$. 

Let \( \bm{Y}_m \) represent the responses corresponding to \( \bm{W}_m \), where \( \bm{Y}_m = (\bm{Y}_g, \bm{Y}_l) \) corresponds to the responses for \( \bm{W}_g \) and \( \bm{W}_l \), respectively.  Although we are unable to use the Wendlands’ compactly supported radial function and fast estimation approach for local points as done in \cite{vakayil2024global}, we can apply the partitioned matrix approach proposed by \cite{barnett1979matrix} to improve the computational efficiency.  
More specifically, the covariance matrix in (\ref{SEzGPcov}) in the SEzGP models can be expressed as
\begin{align}\label{paritionedcovar}
	\boldsymbol{K}_{mm}=\begin{bmatrix}
		\boldsymbol{K}_{gg}& \boldsymbol{K}_{gl}\\
		\boldsymbol{K}_{gl}^\top & \boldsymbol{K}_{ll}
	\end{bmatrix},
\end{align}
\noindent where ${\boldsymbol{K}_{gg}}$ and ${\boldsymbol{K}_{ll}}$ represent the covariance matrix for global points and local points, respectively. 
Applying the partitioned matrix approach, we have
\begin{align}\label{paritionedcovarinverse}
	\boldsymbol{K}_{mm}^{-1}=\begin{bmatrix}
		\boldsymbol{K}_{gg}^{-1}+\boldsymbol{g}_{gl}\boldsymbol{g}_{gl}^\top\boldsymbol{\nu}_{gl}&  \boldsymbol{g}_{gl}\\
		\boldsymbol{g}_{gl}^\top & \boldsymbol{\nu}_{gl}^{-1}
	\end{bmatrix},
\end{align}
where $ {\boldsymbol{g}_{gl}=-\boldsymbol{\nu}_{gl}^{-1}\boldsymbol{K}_{gg}^{-1}\boldsymbol{K}_{gl}}$ and $ {\boldsymbol{\nu}_{gl}=\boldsymbol{K}_{ll}-\boldsymbol{K}_{gl}^\top\boldsymbol{K}_{gg}^{-1}\boldsymbol{K}_{gl}}.$  Using this inverse decomposition, we can quickly update the negative log-likelihood in (\ref{loglikelihood}) such that, 
\begin{align}\label{loglikelihood_n1}
	\begin{split}
		l_{m}(\mu,\bm{\sigma}^2,\bm{\Theta})&= \log|\boldsymbol{K}_{gg}|+\log (\boldsymbol{\nu}_{gl})+(\bm{Y}_{m}-\mu_{m} \bm{\mathds{1}}_m)^\top\boldsymbol{K}_{mm}^{-1}(\bm{Y}_{m}-\mu_{m} \bm{\mathds{1}}_m),
	\end{split}
\end{align} 
where $\mu$, $\bm{\sigma}^2=(\sigma_0^2,\sigma_1^2,\ldots, \sigma_q^2)^\top$, and $\bm{\Theta}=(\bm{\theta}^{(0)},\bm{\theta})$ are the mean, variances, and correlation parameters in (\ref{SEzGPcov}), respectively.  For each predictive location, \( \boldsymbol{K}_{mm}^{-1} \) in (\ref{paritionedcovarinverse}) is updated with the corresponding  local points. The estimate of   \( \mu_m \) is given by,
\begin{align}\label{mun1}
	\hat{\mu}_{m}=(\bm{\mathds{1}}_m^\top\boldsymbol{K}_{mm}^{-1}\bm{\mathds{1}}_m)^{-1}(\bm{\mathds{1}}_m^\top\boldsymbol{K}^{-1}_{mm}\bm{Y}_{m}).
\end{align}
The negative log-likelihood function in (\ref{loglikelihood_n1}) can be minimized given $\hat{\mu}_{m}$ in (\ref{mun1}) to obtain $\hat{\bm{\sigma}}^2$ and $\hat{\bm{\Theta}}$ using an optimization algorithm such as \texttt{rgenoud} in R \citep{r-genoud}, such that,
\begin{align*}
	\medmath{\{\hat{\bm{\sigma}}^2,\hat{\bm{\Theta}}\}=\operatorname*{argmin}_{\bm{\sigma}^2,\bm{\Theta}}\{\log(\boldsymbol{K}_{gg})+\log (\boldsymbol{\nu}_{gl})+(\bm{Y}_{m}^\top\boldsymbol{K}_{mm}^{-1}\bm{Y}_{m})-(\bm{\mathds{1}}_m^\top\boldsymbol{K}_{mm}^{-1}\bm{\mathds{1}}_m)^{-1}(\bm{\mathds{1}}_m^\top\boldsymbol{K}_{mm}^{-1}\bm{Y}_{m})^2\}.}
\end{align*}
For computational efficiency, we parallelize the calculations across predictive locations using the \texttt{foreach} and \texttt{doParallel} packages in R \citep{foreach,doparallel}. Our implementation leverages high-performance computing clusters provided by the Digital Research Alliance of Canada. 
% No OpenMP pragma directives were implemented due to our team's limited expertise   in low-level languages (C/C++/Fortran).  
We call this method as {\em Twin} for the remainder of this article.

\section{Numerical Illustration} \label{sec:numerical-illustration}

In this section, we conduct numerical experiments to evaluate the performance of the methods proposed in Sections 3 and 4. Specifically, we compare the following approaches:
\begin{itemize}
	\item \textbf{SVA}: scaled Vecchia approximation using the SEzGP covariance function;  
	
	\item \textbf{Twin}: an extension of TwinGP with the SEzGP covariance function; 
	\item \textbf{La}: an extension of laGP with the SEzGP covariance function; 
	
	\item \textbf{LE}: the LEzGP model equipped with the SEzGP covariance function; 
	\item \textbf{VA}: non-scaled Vecchia approximation using the SEzGP covariance function;
	\item \textbf{NN}: A nearest-neighbor method combined with the SEzGP model, where neighbors are selected based on the Euclidean distance.
\end{itemize}

All compared methods employ the SEzGP covariance function. For further insight, Appendix A also evaluates their performance using the original EzGP covariance function, further demonstrating the superior performance of the SEzGP covariance function. 

We examine two distinct scenarios:
\begin{itemize}
	\item Scenario 1: Computer experiments with a sufficient number of points for each level combination of the qualitative input variables.
	\item Scenario 2: Computer experiments with a large number of level combinations, resulting from either more qualitative inputs or more levels per input, which leads to fewer design points allocated per level combination for the quantitative inputs.
\end{itemize}

\begin{figure}[htb!]
	\centering
	\includegraphics[scale=0.5]{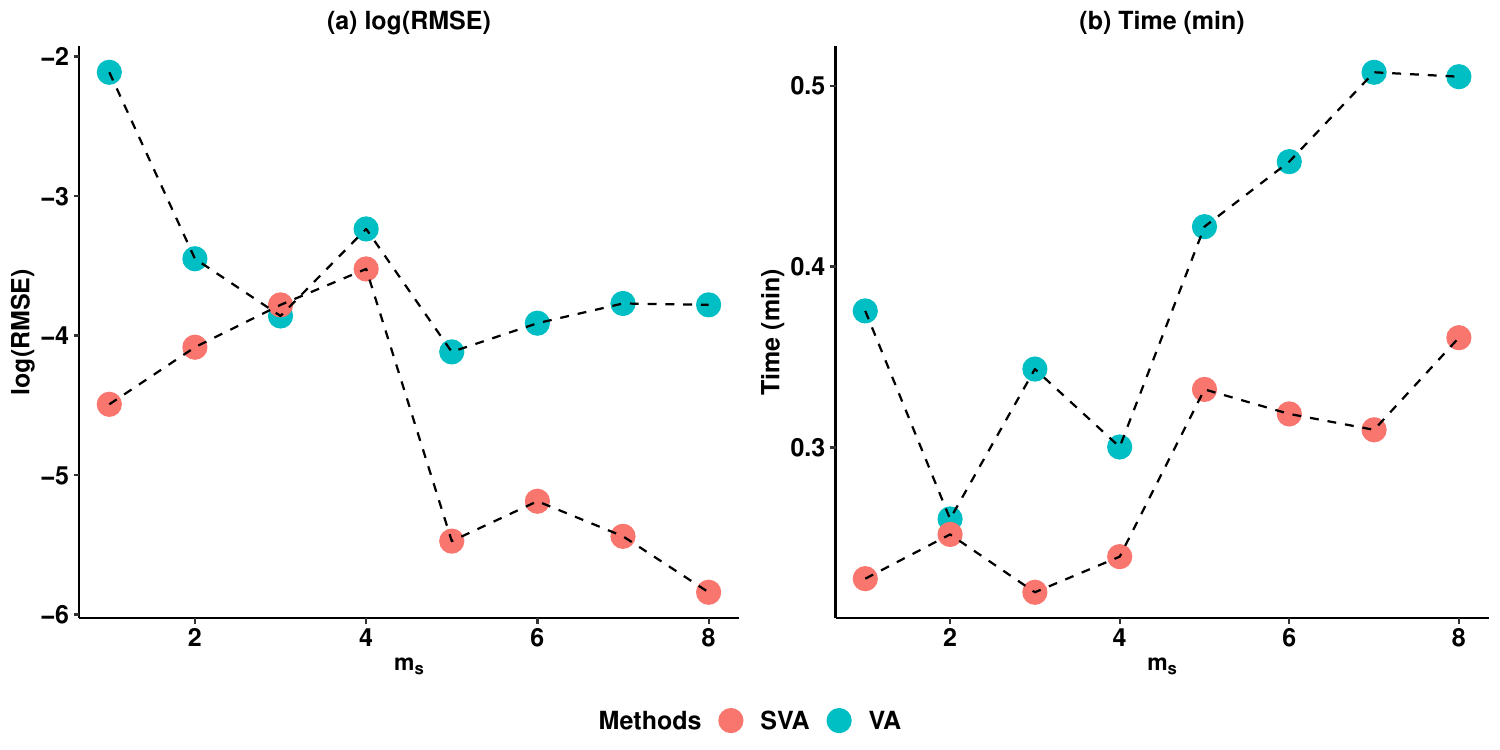}
	\caption{\footnotesize VA and SVA for $m_s=\{1,\ldots,8\}$ in \nameref{ex4.1}: (a) log of RMSE; (b) time in minutes.}\label{Vec_ms}
\end{figure}

Throughout the experiments, we use the following notation: $n$ is the total run size, $p$ and $q$
denote the number of quantitative and qualitative input variables, respectively, and $m$ is the size of the selected subdata. For method-specific parameters, $m_s$ is the conditioning set size for VA, $n_s$ is the tuning parameter in the LEzGP approach, and $g$ and $l$ represent the number of global and local points in Twin. 
As a guideline for tuning parameters, we adopt the sizes for global and local points ($g$ and $l$) recommended by \cite{vakayil2024global} for Twin, and set $m_s = 1$ for $p = 1$ for VA following \cite{katzfuss2022scaled}. After an empirical analysis of the performance, as illustrated in Figure~\ref{Vec_ms} for \nameref{ex4.1}, we observe that the performance improves significantly up to \( m_s = 5 \), with diminishing returns beyond this point.
Thus, we select \( m_s \in \{2, \dots, 5\} \) for \( p > 1 \). 
While these settings may vary with problem specifics and data size, they prove effective for our experimental purposes. Guided by these principles, we establish the following specific configurations for each method:
%  For $p > 1$, we empirically select $m_s \in \{2, \dots, 5\}$.
% Following the recommendations of \cite{vakayil2024global}, we adopt their suggested sizes for global and local points in each method, as outlined below.
% The choice of \( m_s = 1 \) for \( p = 1 \) follows by \cite{katzfuss2022scaled}. For \( p > 1 \), we numerically explore the effect of \( m_s \). As illustrated in Figure~\ref{Vec_ms} for \nameref{ex4.1}, the performance improves significantly up to \( m_s = 5 \), with diminishing returns beyond this point. 

\begin{itemize}
	\item \textbf{Twin}: The size of  global points is set to $g=\min\{50(p+q),\max\{\sqrt{n},10(p+q)\}\}$ and the size of local points is set to $l=\max\{25,3(p+q)\}$. 
	\item \textbf{La}: Starting with an initial set of $l = \max\{25, 3(p+q)\}$ points near $\bm{w}^*$ chosen by the nearest neighbor method,  10 additional points are sequentially added based on the ALC criterion in ~\eqref{ALC_org}.    
	\item \textbf{LE}: The tuning parameter $n_s=q$ considered in the Scenario 1 and $n_s=q-1$ in the  Scenario 2.
	\item \textbf{NN}: The size of local points is set to $l=\max\{25,3(p+q)\}$ plus 10 additional points. 
	\item \textbf{SVA, VA}:  The size of local points  is set to $l=\max\{25,3(p+q)\}$ plus 10 additional points. The conditioning set size  $m_s$  is set to 1 if $p=1$, and between 2 and 5 if $p>1$. 
\end{itemize}

Table~\ref{tab:setup} summarizes the  settings of the methods used in each of four examples.

\begin{table}[htb]
	\centering
	\scalebox{0.8}{
		\begin{tabular}{l l c c c c c}
			\hline
			Example & Scenario & $n$ & Twin $(g,l,m)$ & NN/La $(l,m)$ & VA/SVA $(l,m,m_s)$ & LE $(m,n_s)$ \\
			\hline
			Examples 1--2 & Scenario 1 & 5400 
			& $(80,25,105)$ 
			& $(25,35)$ 
			& $(25,35,5)$ 
			& $(200,3)$ \\
			
			Examples 1--2 & Scenario 2 & 5000 
			& $(73,25,98)$ 
			& $(25,35)$ 
			& $(25,35,5)$ 
			& $(140,2)$ \\
			
			Example 3 & Scenario 1 & 5400 
			& $(73,25,98)$ 
			& $(25,35)$ 
			& $(25,35,3)$ 
			& $(200,3)$ \\
			
			Example 4 & Scenario 2 & 3645 
			& $(71,25,96)$ 
			& $(25,35)$ 
			& $(25,35,1)$ 
			& $(140,5)$ \\
			\hline
	\end{tabular}}
	\caption{Summary of experimental configurations for each simulation scenario, with the total training points and the respective tuning parameters.}
	\label{tab:setup}
\end{table}

% one-hot coding is sort of standard, and "retaining the original data" was mentioned in line 355. (right above eqn 4.6), so I deleted the below.
% For one-hot encoding, we use the \texttt{one\_hot} function from the \texttt{mltools} package in R \citep{r-onehot}. While the subset selection is based on the distances after applying one-hot encoding, the original data is retained for model fitting and prediction. Thus, this approach is solely used to select local and global points from the full dataset.

The training data are generated as follows. A random Latin hypercube design (using the \texttt{randomLHS} function in the \texttt{lhs} package in R \citep{lhs-package})  is used for choosing quantitative inputs, while qualitative inputs include all level combinations. To evaluate the performance of different methods, we compare the predicted responses and the true responses using the root mean square error (RMSE):
\begin{align}\label{RMSE}
	\text{RMSE} &= \sqrt{\frac{1}{n_w}\sum_{i=1}^{n_w} \left(\hat{Y}(\bm{w}^{*}_i) - Y(\bm{w}^{*}_i)\right)^2},
\end{align}
where $n_w$ denotes the number of test points. The test set is constructed by first generating $n_w$ points for quantitative inputs at each qualitative level combination via \texttt{randomLHS} \citep{lhs-package} and then randomly selecting $n_w$ test locations from these points. 
This approach ensures random test location selection while maintaining a controlled test set size for fair comparison.
In the captions of the figures in the examples, the values in the parenthesis represent the total size of the subdata in each method, and in the figures, we present the logarithm of the RMSE (base \( e \)).  All the implementations can be found at \url{https://github.com/devonlin/Emulator_for_large_mixed_inputs_computer_experiments}.

\begin{comment}
	\begin{table}[thb]
		\centering
		\begin{tabular}{l l  l l l l l l l}  
			\hline
			& &   $n$ &  Method & $g$ & $l$ & $m$ & $m_s$ & $n_s$\\
			\hline
			Examples 1 and 2  & Scenario 1 &  5400  & Twin & 80 & 25 & 105 &  & \\
			& & & NN, La  & & 25 & 35 & &  \\
			& & &  VA, SVA & & 25 & 35 & 5 & \\
			& & &  LE & &  & 200 &   & 3\\
			& Scenario 2 &  5000  & Twin & 73 & 25 & 98 &  & \\
			& & & NN, La  & & 25 & 35 & &  \\
			& & &  VA, SVA & & 25 & 35 & 5 & \\
			& & &  LE & &  & 140 &   & 2\\
			&&&&&&&&\\
			Example 3 & Scenario 1 &  5400  & Twin & 73 & 25 & 98 &  & \\
			& & & NN, La  & & 25 & 35 & &  \\
			& & &  VA, SVA & & 25 & 35 & 3 & \\
			& & &  LE & &  & 200 &   & 3\\
			&&&&&&&&\\
			Example 4& Scenario 2 & 3645  & Twin & 71 & 25 & 96 &  & \\
			& & & NN, La  & & 25 & 35 & &  \\
			& & &  VA, SVA & & 25 & 35 & 1 & \\
			& & &  LE & &  & 140 &   & 5\\
			\hline 
		\end{tabular}
		\caption{Setup of the methods}
		\label{tab:setup}
	\end{table}
\end{comment}

\subsection*{Example 1}\label{ex4.1}
We consider a computer model with $p=4$ quantitative inputs $\bm{x}=(x_1,x_2,x_3,x_4)$ and $q=3$ qualitative inputs $\bm{z}=(z_1,z_2,z_3)$. The computer model is  represented by, 
\begin{align*}
	f(\bm{x},\bm{z})=\frac{2\pi x_1z_1}{\log(z_2/x_2)(1.5+\frac{2x_3x_1}{\log(z_2/x_2)x_2^2z_3}+\frac{x_1}{x_4})}.
\end{align*}
\noindent A similar function was considered by \cite{zhou2011simple}. The comparative setup and results for Scenarios 1 and 2 are discussed below.

\begin{itemize}
	\item {Scenario 1}\\
	We consider three levels for each qualitative input, yielding \(3^3 = 27\) level combinations. The training data consists of 200 points per level combination and thus we have \(200 \times 27 = 5400\) points in training data.   The settings of all methods are summarized in Table~\ref{tab:setup}.
	%The methods use the following configurations:
	%\begin{itemize}
	%    \item \textbf{Twin}: \(g = 80\) global points and \(l = 25\) %local points (\(m = 105\) total)
	%    \item \textbf{NN, La, VA, and SVA}: \(l = 25\) local points plus %10 additional points (\(m = 35\) total)
	%    \item \textbf{SVA and VA}: Conditioning set size \(m_s = 5\)
	%    \item \textbf{LE}: Tuning parameter \(n_s = 3\)
	%\end{itemize}
	Figure~\ref{box_Ex1_1} shows the log-RMSE and average computation time (in minutes) across 30 simulations for all methods using the SEzGP covariance function in (\ref{SEzGPcov}). 
	It also reveals that SVA achieves the lowest RMSE, demonstrating superior prediction accuracy among all methods. While NN offers the fastest computation time, it comes at the cost of significantly worse accuracy. Importantly, SVA maintains competitive computation times while delivering substantially better predictive performance, striking a balance between these two crucial metrics, making SVA the preferred choice.
	
	\begin{figure}[htb!]
		\centering
		\includegraphics[scale=0.5]{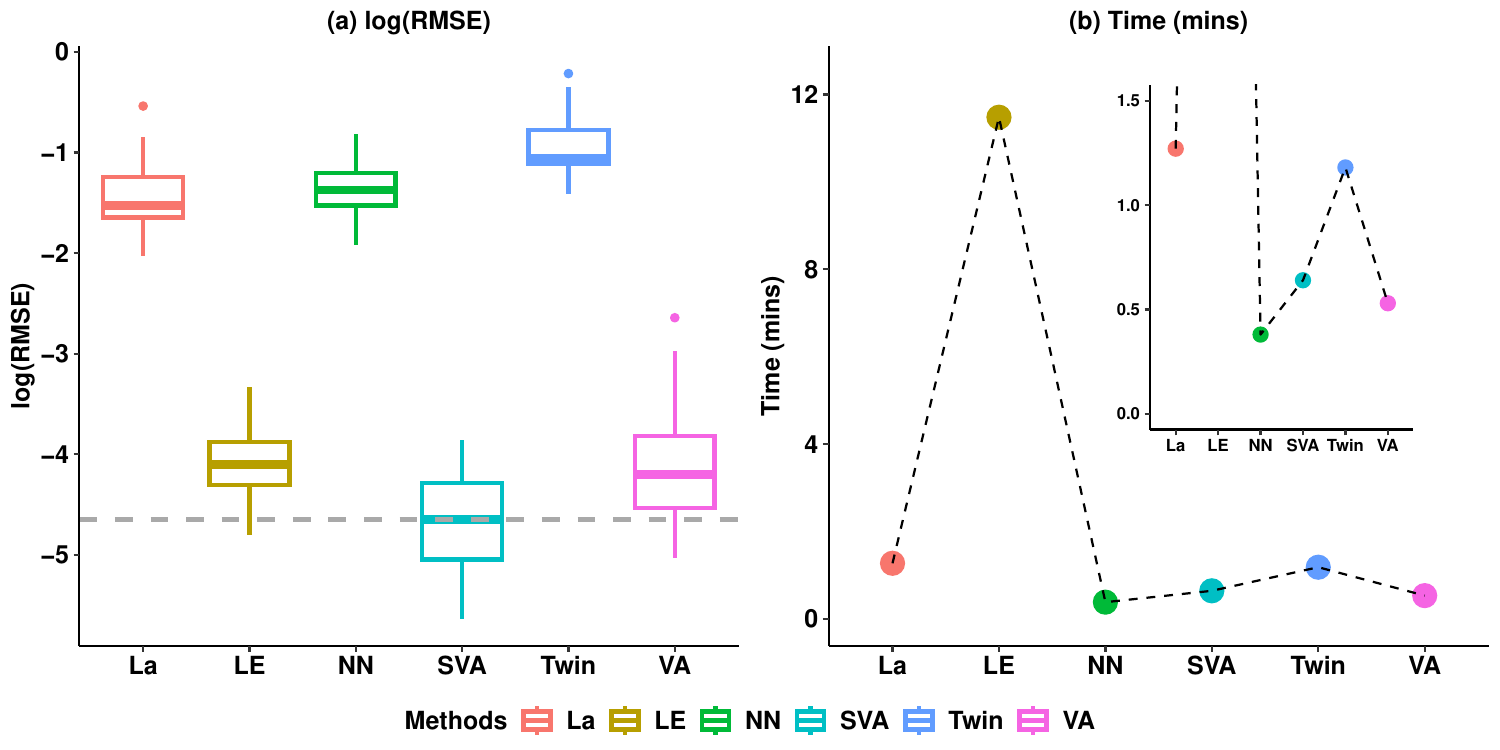}
		\caption{\footnotesize  Simulation results for  \nameref{ex4.1}, Scenario 1 across 30 replications:    (a) The boxplot of log-RMSE; (b)  The average of time in minutes, with Twin (105), LE (200), NN (35), La (35), VA (35) and SVA (35). The inset shows a zoomed-in view of plot (b).}\label{box_Ex1_1}
	\end{figure}
	
	\item {Scenario 2}\\
	We consider ten levels for each qualitative input variable, generating \(10^3 = 1000\) level combinations. The training dataset contains 5 points per level combination, totaling \(5 \times 1000 = 5000\) points.  The settings of all methods can be found in Table~\ref{tab:setup}.
	
	%The methods employ the following configurations:
	%\begin{itemize}
	%    \item \textbf{Twin}: \(g = 73\) global points and \(l = 25\) %local points (\(m = 98\) total)
	%    \item \textbf{NN, La, VA, and SVA}: \(l = 25\) local points plus %10 additional points (\(m = 35\) total)
	%    \item \textbf{SVA and VA}: Conditioning set size \(m_s = 5\)
	%    \item \textbf{LE}: Tuning parameter \(n_s = 2\)
	%\end{itemize}

	\begin{figure}[htb!]
		\centering
		\includegraphics[scale=0.5]{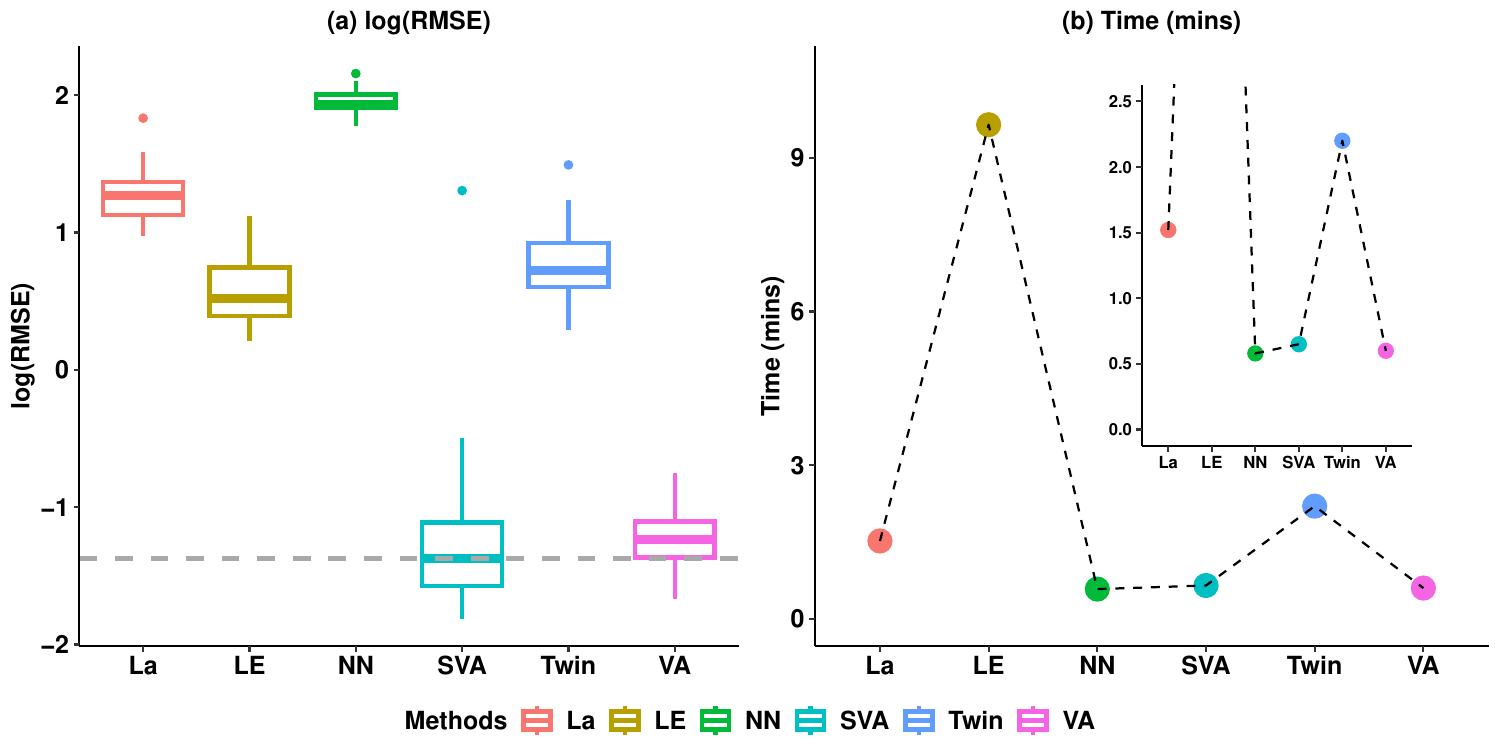}
		\caption{\footnotesize  Simulation results for  \nameref{ex4.1}, Scenario 2 across 30 replications:  (a) The boxplot of log-RMSE; (b) The average of time in minutes, with Twin (98), LE (140), NN (35), La (35), VA (35) and SVA (35).}\label{box_Ex2_1}
	\end{figure}
	
	Following the same approach as Scenario 1, Figure~\ref{box_Ex2_1} displays the logarithm of the RMSE and average computation time (in minutes) across 30 simulations using the SEzGP covariance function in (\ref{SEzGPcov}) in Scenario 2. The results demonstrate that SVA achieves superior prediction accuracy with the lowest RMSE values among all methods. While NN takes least computation time, its accuracy is significantly reduced. Again, SVA offers computational efficiency comparable to NN while delivering substantially better predictive performance. This superior accuracy-efficiency trade-off establishes SVA as the most practical and effective choice among the evaluated methods.
	
\end{itemize}

%!!!!!!!!!!!!!!!!!!!!!!!!!!!!!!!!!!!!!!!!!!!!!!!!!!!!!!!!!!!!!!!!!!!!!!!!!!!!!!!!!!
%!!!!!!!!!!!!!!!!!!!!!!!!!!!!!!!!!!!!!!!!!!!!!!!!!!!!!!!!!!!!!!!!!!!!!!!!!!!!!!!!!!!
\subsection*{Example 2}\label{ex4.2}
We consider a computer model with 8 input variables  in which, the five input variables $x_1,\ldots,x_5$ are quantitative  and the three inputs $z_1,z_2,z_3$ are qualitative. The computer model \citep{an2001quasi} is represented by, 
\begin{align*}
	\begin{split}
		&u=\sum_{i=1}^{4}L_i\cos(\sum_{j=1}^{i}\theta_j)\\
		&v=\sum_{i=1}^{4}L_i\sin(\sum_{j=1}^{i}\theta_j)\\
		&f(\bm{w})=10(u^2+v^2)^{0.5},
	\end{split}
\end{align*}
where $\bm{w}=(\theta_1, \theta_2,\ldots, L_3, L_4)^\top$ with $\theta_1 = x_1, \theta_2 = x_2, \theta_3 = x_3, \theta_4 = x_4, L_1 = x_5, L_2 = z_1,   L_3=z_2, L_4=z_3$.
The setup and results for Scenarios 1 and 2 are as follows.

\begin{itemize}
	\item{{Scenario 1}}
	
	We consider three levels for each qualitative input, resulting  \(3^3 = 27\) level combinations. The training dataset consists of 200 points per level combination, resulting in a total of \(200 \times 3^3 = 5400\) points.  
	%The settings of all methods can be found in Table~\ref{tab:setup}.
	Figure~\ref{box_Ex1_22} shows the log-RMSE and average computation time (in minutes) over 30 simulation runs for the six methods using the SEzGP covariance function  in (\ref{SEzGPcov}) under Scenario~1. Figure~\ref{box_Ex1_22} reveals that the  SVA method consistently achieves the lowest RMSE values and shorter computation times compared to the other methods, making it the most effective approach among the competitors.
	%For the Twin method, we use \(g = 80\) global points and \(l = 25\) local points, yielding a total of \(m = l + g = 105\) points. For the NN,  La,  VA, and  SVA methods, we set \(l = 25\) and include 10 additional points, resulting \(l + 10 = 35\) points. In the SVA and  VA methods, we also set \(m_s = 5\), while in the  LE method, we use \(n_s = 3\).

	\begin{figure}[htb!]
		\centering
		\includegraphics[scale=0.5]{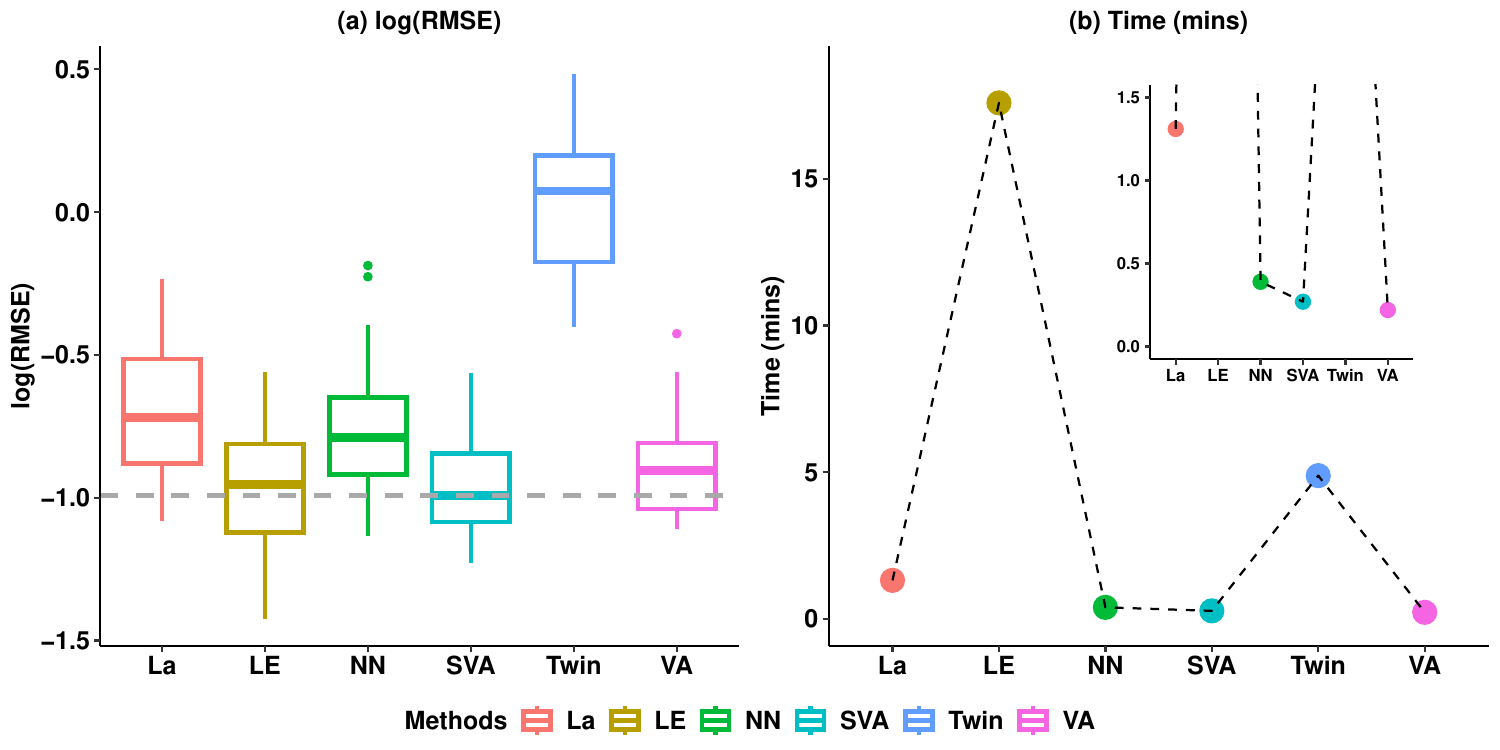}
		\caption{\footnotesize Simulation results for \nameref{ex4.2} under Scenario 1 across 30 replications: (a) The boxplot of log-RMSE; (b)  The average of time in minutes per simulation, with Twin (105), LE (200), NN (35), La (35), VA (35) and SVA (35).}\label{box_Ex1_22}
	\end{figure}
	
	\item {Scenario 2}
	
	We consider ten levels for each qualitative input variable, yielding \(10^3 = 1000\) level combinations. The training dataset contains 5 points per level combination, totalling \(5 \times 10^3 = 5000\) points.   
	%The settings of all methods are summarized in Table~\ref{tab:setup}.
	%For the Twin  method, we use \(g = 80\) global points and \(l = 25\) local points, giving a total of \(m = l + g = 105\) points. For the NN, La, VA, and  SVA methods, we set \(l = 25\) local points and include 10 additional points, resulting \(l + 10 = 35\) total points. In the  SVA and VA methods, we  specify \(m_s = 5\), while for the LE method, we use \(n_s = 2\).
	Figure~\ref{box_Ex2_2} presents the log-RMSE and the average computation time (in minutes) across 30 simulations for the six methods, using the SEzGP covariance function in (\ref{SEzGPcov}) under Scenario~2. Figure~\ref{box_Ex2_2} reveals that while LE method achieves the lowest RMSE values, it incurs the highest computational cost. In contrast,  SVA provides the second-best RMSE performance with substantially lower computation time. This shows that SVA offers a more favorable trade-off between accuracy and efficiency.

	\begin{figure}[htb!]
		\centering
		\includegraphics[scale=0.5]{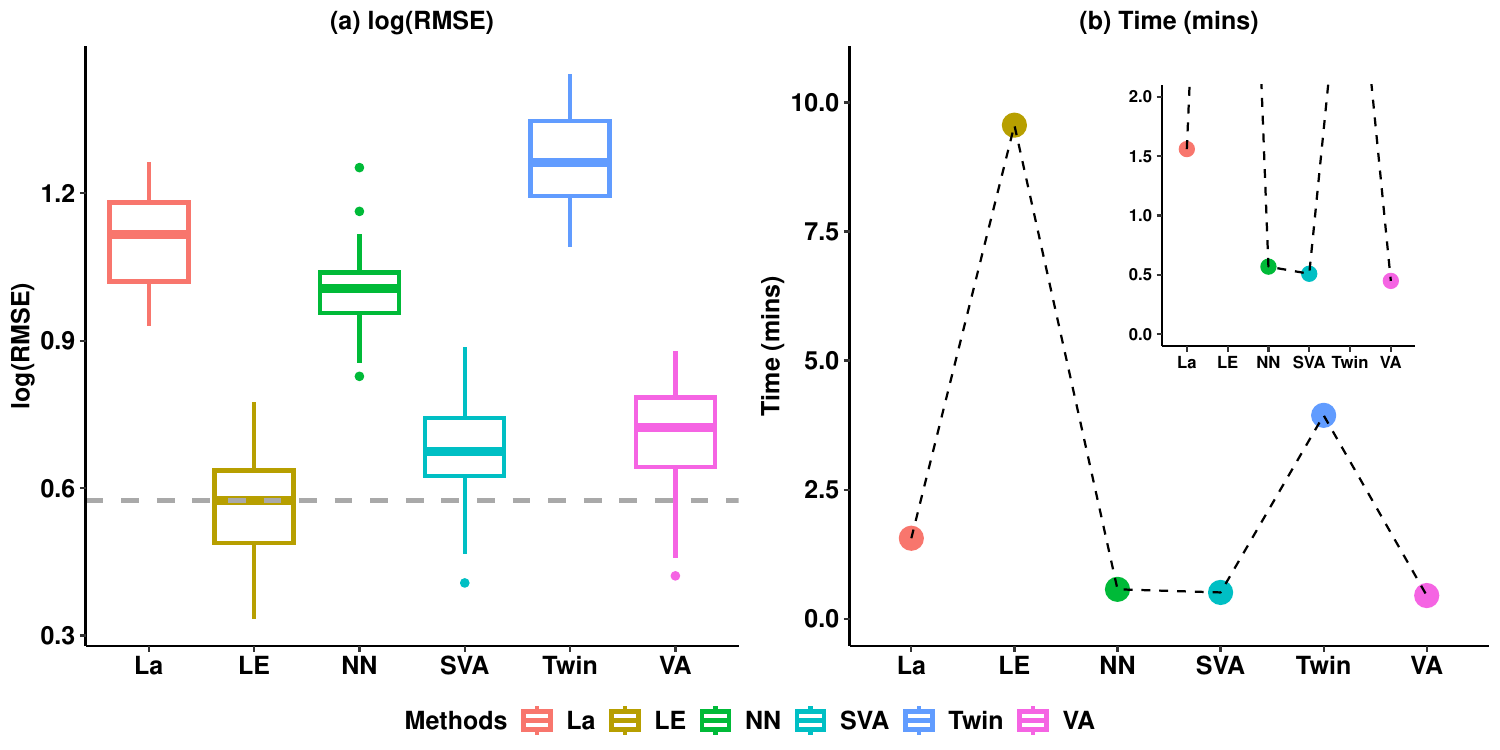}
		\caption{\footnotesize Simulation results for \nameref{ex4.2}, Scenario 2 across 30 replications:   (a) The boxplot of log-RMSE; (b)  The average of time in minutes, with Twin (105), LE (140), NN (35), La (35), VA (35) and SVA (35).}\label{box_Ex2_2}
	\end{figure}
\end{itemize}

\subsection*{Example 3}\label{ex4.3}
We consider a computer model with $p=3$ quantitative input variables $\bm{x}=(x_1,\ldots,x_3)$ and $q=3$ qualitative input variables $\bm{z}=(z_1,z_2,z_3)$. Following  \citep{xiao2021ezgp}, the function is defined as
{\footnotesize
	\begin{align*}
		i(\bm{x},{z}_1)=
		\begin{cases}
			x_1+x_2^2+x_3, & \text{if}\ z_1=1 \\
			x_1^2+x_2+x_3, & \text{if}\ z_1=2\\
			x_3+x_1+x_2^2,&  \text{if}\ z_1=3,
		\end{cases}~~
		g(\bm{x},{z}_2)=
		\begin{cases}
			\cos( x_1)+\cos(2 x_2)+\cos(x_3), & \text{if}\ z_2=1 \\
			\cos( x_1)+\cos(2 x_2)+\cos(x_3), & \text{if}\ z_2=2\\
			\cos(2x_1)+\cos(x_2)+\cos(x_3),&  \text{if}\ z_2=3,
		\end{cases}
	\end{align*}
	\begin{align*}
		h(\bm{x},{z}_3)=
		\begin{cases}
			\sin(x_1)+\sin(2 x_2)+\sin(x_3), & \text{if}\ z_3=1 \\
			\sin(x_1)+\sin(2 x_2)+\sin(x_3), & \text{if}\ z_3=2\\
			\sin(2x_1)+\sin(x_2)+\sin(x_3),&  \text{if}\ z_3=3,
		\end{cases}~~
		f(\bm{x},\bm{z})=100 i(\bm{x},z_1)+g(\bm{x},z_2)+h(\bm{x},z_3).
	\end{align*}
}

Due to the structure of this specific test function, increasing the number of levels for each qualitative variable to 10, as done in \nameref{ex4.1} and \nameref{ex4.2},
is not feasible. As such, we limit our comparison for this example for Scenario 1. 
% In \nameref{ex4.3}, if we use five data points for each level combination of the qualitative input variables, the total number of data points is \(5 \times 3^3 = 135\), which is insufficient to represent a large-scale computer experiment.

% Below, we present the comparison framework and results for Scenario 1, and discuss the reason for not considering  Scenario 2.

%\begin{itemize}
%\item {Scenario 1}

\begin{figure}[htb!]
	\centering
	\centering
	\includegraphics[scale=0.5]{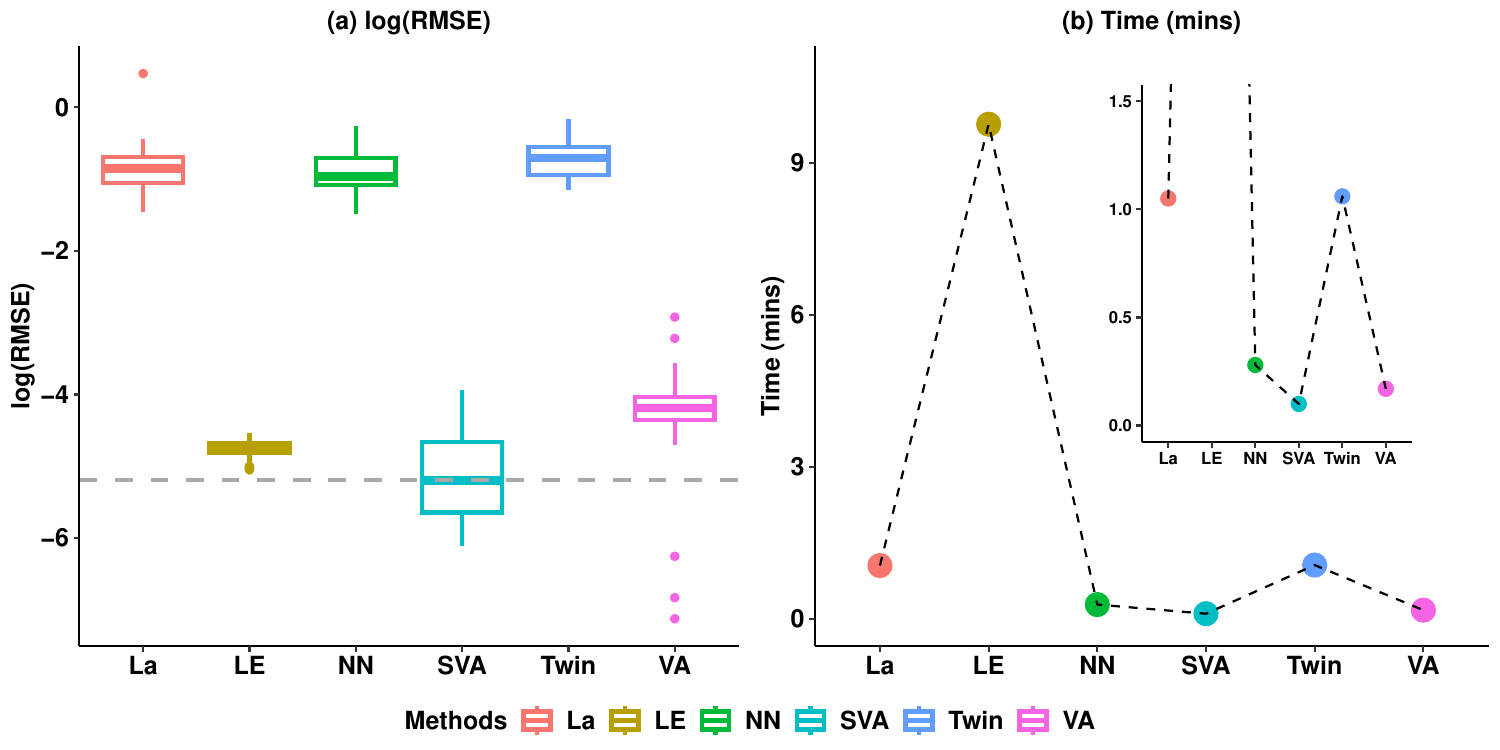}
	\caption{ {\footnotesize Simulation results for \nameref{ex4.3} under Scenario 1 across 30 replications:  (a) The boxplot of RMSE; (b) The average of time in minutes, with Twin (98), LE (200), NN (35), La (35), VA (35) and SVA (35).}}\label{box_Ex1_3}
\end{figure}

We consider three levels for each qualitative input and thus we have  \(3^3 = 27\) level combinations in total. The training data consists of 200 points for each level combination, resulting in a total of \(200 \times 3^3 = 5400\) points. 
% See Table~\ref{tab:setup} for the settings of all methods.
%For the Twin method,  we set \(g = 73\) for the global inputs, and for the local inputs, we use \(l = 25\) points per simulation, leading to a total of \(m = l + g = 98\) input dimensions. For the NN, La, VA, and SVA methods, the number of local points is \(l = 25\), augmented by 10 additional points, resulting in \(l + 10 = 35\) local points. In the SVA and VA methods, we set \(m_s = 3\), while in the LE method, we use \(n_s = 3\).
Figure~\ref{box_Ex1_3} displays the log-RMSE and the average computation time (in minutes) across 30 simulations for the evaluated methods. The figure indicates that the SVA method consistently outperforms the other approaches in both prediction accuracy and computational efficiency. Specifically, SVA achieves lower RMSE values while requiring less computational time.

% Due to the nature of the function, it is not feasible to increase the number of levels for each qualitative variable to 10, as was done in \nameref{ex4.1} and \nameref{ex4.2}. In \nameref{ex4.3}, if we use five data points for each level combination of the qualitative input variables, the total number of data points is \(5 \times 3^3 = 135\), which is insufficient to represent a large-scale computer experiment.
% To better illustrate the expected behavior of such computer simulators, we present \nameref{ex4.4} under Scenario 2. This example uses a similar class of functions but incorporates additional qualitative factors, enabling a more representative demonstration of the results in a larger-scale setting.

%\end{itemize}
%!!!!!!!!!!!!!!!!!!!!!!!!!!!!!!!!!!!!!!!!!!!!!!!!!!!!!!!!!!!!!!!!!!!!!!!!!!!!!!!!!!
%!!!!!!!!!!!!!!!!!!!!!!!!!!!!!!!!!!!!!!!!!!!!!!!!!!!!!!!!!!!!!!!!!!!!!!!!!!!!!!!!!!!

\subsection*{Example 4}\label{ex4.4}
This example is designed to complement the results from \nameref{ex4.3}. 
This setup utilizes a similar class of functions but incorporates additional qualitative factors to create a larger-scale setting, 
to provide a more representative demonstration of the results when design points per level combination are limited.
We consider a computer model with $p=1$ quantitative input $x$ and $q=6$ qualitative inputs $\bm{z}=(z_1,z_2,z_3,z_4,z_5,z_6)$. 
The computer model is defined as

{
	\begin{align*}
		f(x,\bm{z})=h(z_6)+e(z_5)+\left[\left\{(a(x,z_1)+b(x,z_2)\right\}\cdot c(x,z_3) \cdot d(z_4)\right],
	\end{align*}
}

\noindent where the functions $\{a,b,c,d,e,h\}$ are given by

{ \footnotesize
	\begin{align*}
		a(x,{z}_1)=
		\begin{cases}
			\cos(3\pi x ),& \text{if}\ z_1=1 \\
			\cos(4\pi x ), & \text{if}\ z_1=2\\
			\cos(5\pi x ), & \text{if}\ z_1=3,\\
		\end{cases}~~~~
		b(x,{z}_2)=
		\begin{cases}
			\sin(3\pi x ),& \text{if}\ z_2=1 \\
			\sin(4\pi x), & \text{if}\ z_2=2\\
			\sin(5\pi x), & \text{if}\ z_2=3,\\
		\end{cases}~~~~
		c({z}_3)=
		\begin{cases}
			1, & \text{if}\ z_3=1 \\
			2, & \text{if}\ z_3=2\\
			3, & \text{if}\ z_3=3,\\
		\end{cases}
	\end{align*}
	\begin{align*}
		d({z}_4)=
		\begin{cases}
			0.1, & \text{if}\ z_4=1 \\
			0.2, & \text{if}\ z_4=2\\
			0.3, & \text{if}\ z_4=3,\\
		\end{cases}~~~~     
		e(x,{z}_5)=
		\begin{cases}
			x, & \text{if}\ z_5=1 \\
			x^2, & \text{if}\ z_5=2\\
			x^3, & \text{if}\ z_5=3,\\
		\end{cases}~~~~     
		h({z}_6)=
		\begin{cases}
			-2, & \text{if}\ z_6=1 \\
			0, & \text{if}\ z_6=2\\
			2, & \text{if}\ z_6=3.\\
		\end{cases}    \\
	\end{align*}
}

\begin{figure}[htb!]
	\centering
	\centering
	\includegraphics[scale=0.5]{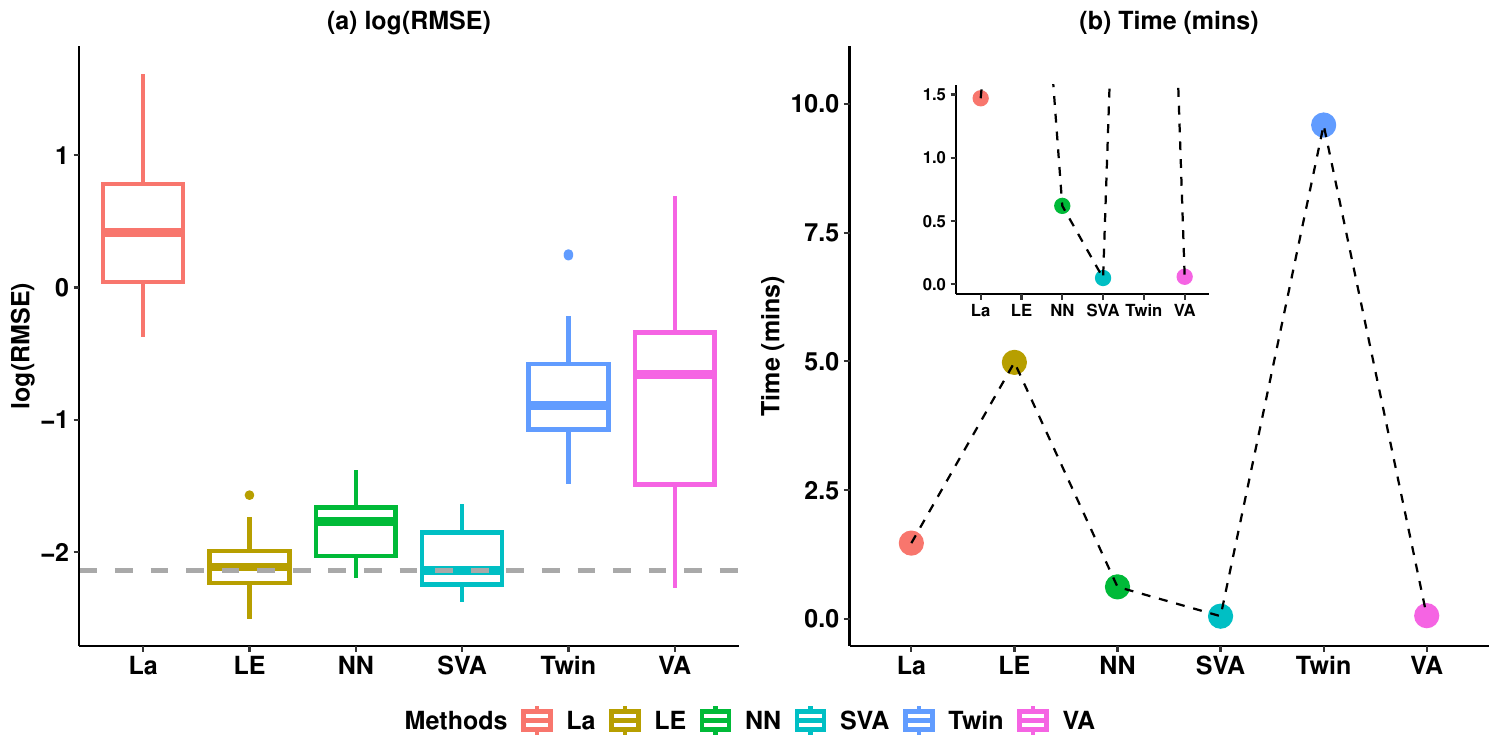}
	\caption{\footnotesize Simulation results for \nameref{ex4.4} under Scenario 2 across 30 replications:  (a) The boxplot of RMSE; (b) The average of time in minutes, with Twin (96), LE (65), NN (35), La (35), VA (35) and SVA (35).}\label{box_Ex2_3}
\end{figure}

We now present the comparison framework and results for Scenario 2.
%\begin{itemize}
%\item {Scenario 2}
We consider three levels for each qualitative input and thus there are a total of \(3^6 = 729\) level combinations. The training data consists of 5 points 
per level combination, 
resulting in a total of \(5 \times 3^6 = 3645\) data points.  
% The settings of all methods are summarized in Table~\ref{tab:setup}.
%For the Twin method, we set \(g = 71\) for the global inputs and \(l = 25\) for the local inputs, leading to a total of \(m = l + g = 96\) points. For the NN, La, VA, and SVA methods, the number of local points is \(l = 25\), augmented by 10 additional points, yielding in \(l + 10 = 35\) local points. In the SVA and VA methods, we set \(m_s = 1\), while in the LE method, we consider \(n_s = 5\).
Figure~\ref{box_Ex2_3} displays the log-RMSE and the average computation time (in minutes) across 30 simulations for the evaluated methods. 
Figure~\ref{box_Ex2_3} reveals that the SVA method consistently outperforms the other approaches in both prediction accuracy and computational efficiency. Specifically, SVA achieves lower RMSE values, %indicating more accurate predictions, while also requiring less computational time, 
making it the most efficient method overall. It is worth pointing out the SVA significantly outperforms the VA in this example, highlighting the
advantage of scaled framework.

%\end{itemize}

\section{A Real Application} \label{sec:case-study}
In this section, we present a case study to evaluate the predictive and computational performance of the methods, motivated by an engineering application. Specifically, we focus on beam deflection, a primary response metric used in
structural engineering to prevent cracking and ensure overall system stability. 
To create a realistic dataset, we synthesized beam deflection cases by incorporating the geometries from the  American Institute of Steel Construction (AISC) shapes database  \citep{AISC2013}. 
The continuous inputs involved in the physical process are the depth and the cross-sectional area of the beam, the span length, and the load position along the beam. This simulation is based on a physics model that calculates the maximum deflection of a simply-supported beam subjected to a point load of unit magnitude.  The magnitude of deflection is governed by the beam's physical geometric properties, material properties (i.e., the modulus of elasticity), and the location of the applied load.

To ensure the synthesized cases strictly follow physical constraints and simulate realistic behaviors, 
we used the W-shape (H-beam) geometric configurations from the AISC database as a baseline. 
For the comparative study, dimensions for other cross-sections, such as circular, 
rectangular shapes, and T-sections, were synthesized to be physically comparable 
to the actual geometries found in the AISC database. 
For each geometric configuration, values for the remaining quantitative inputs, span length and load position ratio, 
were sampled using a maximin Latin hypercube design. 
Two qualitative inputs are materials (four levels: steel, aluminum, concrete and wood) and cross sections (five levels: rectangular shape with two width-to-height ratios, circular, T-shapes, and H-shapes).
The complete specification of the input parameters and the response variable is summarized in Table \ref{tab:variable-summary}.
The training data consist of 8,000 runs based on 20 geometric conditions and 20 combinations of length and load positions across all five shapes and four materials. Similarly, the test data consist of 2,000 runs generated from 100 geometric conditions and one combination of length and load positions for each of 30 simulation replications.
%Similarly, the test data includes 200,000 runs, 
%comprising 100 geometric conditions, 
%100 combinations of length and load positions.
In this case study, the LE method was excluded from the comparison due to its prohibitive computational cost.
% \textcolor{teal}{I used the numbers I provided, but it seems you used 30 cases? }
\begin{table}[thb]
	\centering
	\scalebox{0.85}{
		\begin{tabular}{lll} \hline
			Variable Type &  Variables        & Description \\ \hline
			Quantitative  &  Beam Length      & Maximin design  \\
			&  Load Ratio       & Maximin design \\ 
			&  Area             & AISC Database \\
			&  Depth            & AISC Database \\  \hline
			Qualitative   &  Material         & Steel, Aluminum, Concrete, Wood \\
			&  Shape            & H-shape, T-shape, Circular, Rect (1:2), Rect (1:4) \\ \hline
			Response      &  Maximum Deflection & Simulated via physics-based model \\ \hline
	\end{tabular}}
	\caption{Variable specifications for the large-scale beam deflection case study, detailing both the input parameters and the response variable.}
	\label{tab:variable-summary}
\end{table}

\begin{figure}[htb!]
	\centering
	\centering
	\includegraphics[scale=0.5]{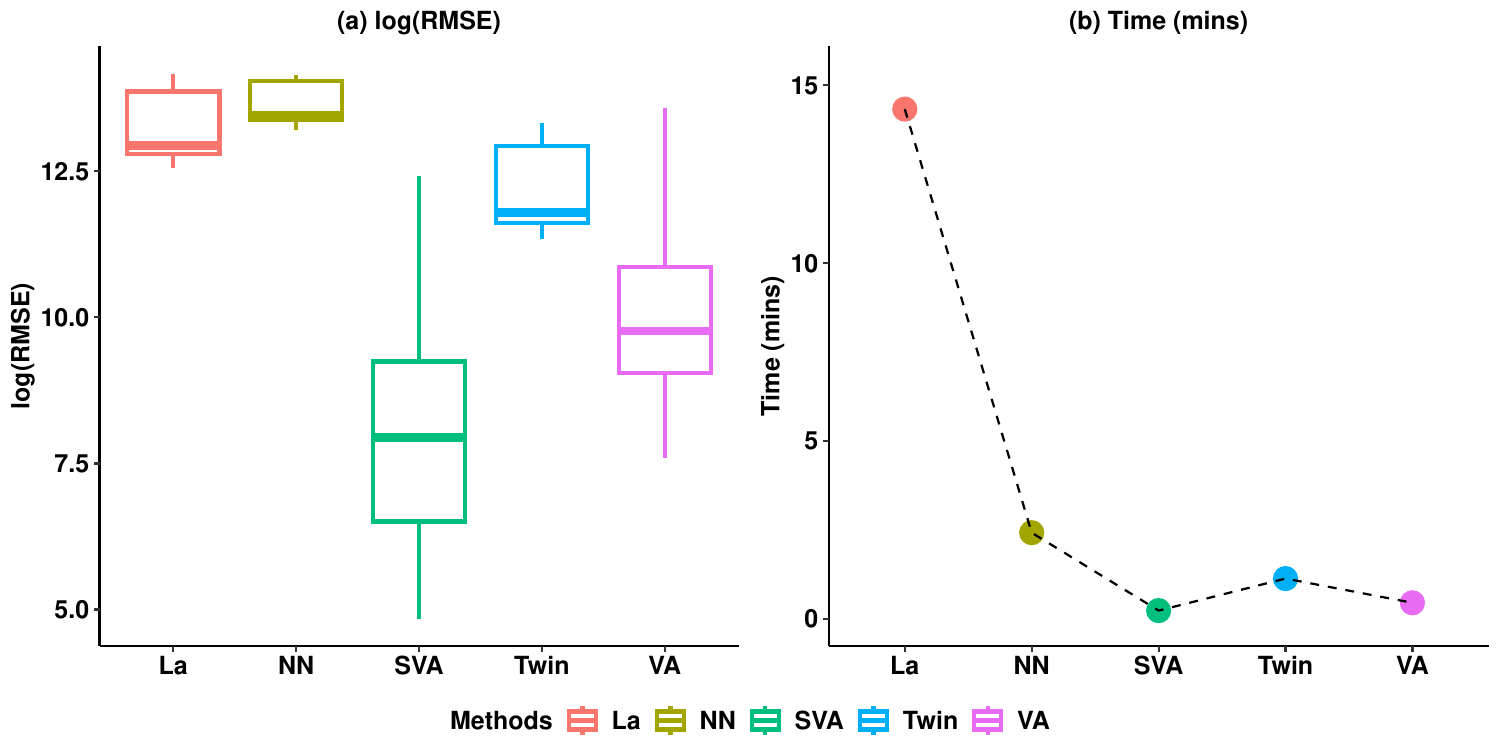}
	\caption{\footnotesize Performance comparison across 30 simulations for the case study: (a) boxplots of log-RMSE; (b) average computation time in minutes. The values in parentheses represent the number of points used for modeling for each method: Twin (89), NN (25), La (35), VA (35), and SVA (35).}\label{fig:real}
\end{figure}
% \textcolor{teal}{What are the numbers like (89) in the caption?}
Figure~\ref{fig:real} presents the log-RMSE and the average computation time (in minutes) across 30 simulations for the five methods. As shown in Figure~\ref{fig:real}, the SVA method generally outperforms the other approaches in both predictive accuracy and computational efficiency.
%, although the performance gains are less pronounced in this case study.  

\section{Concluding Remarks} \label{sec:conclusion}

In this work, we investigated several techniques for large-scale computer experiments involving both quantitative and qualitative inputs. In particular, we proposed a novel scaled Vecchia approximation method, termed SEzGP, for such experiments, as described in Section~\ref{sec:covariance-function}. We also extended two prominent methods for large-scale computer experiments with quantitative inputs, laGP and TwinGP, to accommodate mixed-input settings.

Vecchia approximations employ an ordered conditional representation that enables efficient joint likelihood estimation, prediction, and uncertainty quantification. The proposed SEzGP method incorporates scaling parameters that automatically capture the effects of both quantitative and qualitative inputs through a fast parameter estimation procedure. By leveraging the advantages of Vecchia approximations, the proposed approach improves both predictive accuracy and computational efficiency for computer experiments with mixed inputs. Numerical studies and a real-data application demonstrate that the proposed method consistently outperforms competing approaches, including La, Twin, NN, LE, and VA, across a variety of large-scale computer experiment settings. Although our extensions of laGP and TwinGP did not exhibit the same advantages as observed in large-scale computer experiments with only quantitative inputs, more sophisticated extensions may further reveal their potential for the mixed-input settings.

In this study, the EzGP framework was used to construct the surrogate model. Other frameworks for mixed inputs, such as LvGP \citep{zhang2020latent} and category tree GP \citep{lin2024category}, are also  worth investigating in the context of large-scale computer experiments.  For future research, we plan to investigate the use of the proposed covariance function in inference tasks, such as global optimization and  contour estimation for large-scale computer experiments with mixed inputs.

\section*{Acknowledgment}

Lin was supported by a Discovery grant from the Natural Sciences and Engineering
Research Council of Canada.
Part of this research was performed while  Hwang and Lin were  visiting the
Institute for Mathematical and Statistical Innovation (IMSI) at University of Chicago from March 3 to May 24, 2025, which is supported by the National Science Foundation (Grant No. DMS-1929348). We also thank Digital Research Alliance of Canada for providing clusters for computing. 

%\appendix

\appendix
\section*{Appendix}
\renewcommand{\thesubsection}{\Alph{subsection}}

\subsection{Comparison of EzGP and SEzGP}

We compare the prediction accuracy and computational efficiency of different methods for large-scale computer experiments involving both quantitative and qualitative inputs. In particular, we focus on the EzGP and SEzGP covariance functions. The motivation for these comparisons is to highlight the performance advantages of the SEzGP covariance function, defined in~(\ref{SEzGPcov}), over the standard EzGP covariance function.

\begin{figure}[!htb]
	\centering
	\begin{subfigure}[b]{\textwidth}
		\centering
		\includegraphics[width=0.85\textwidth]{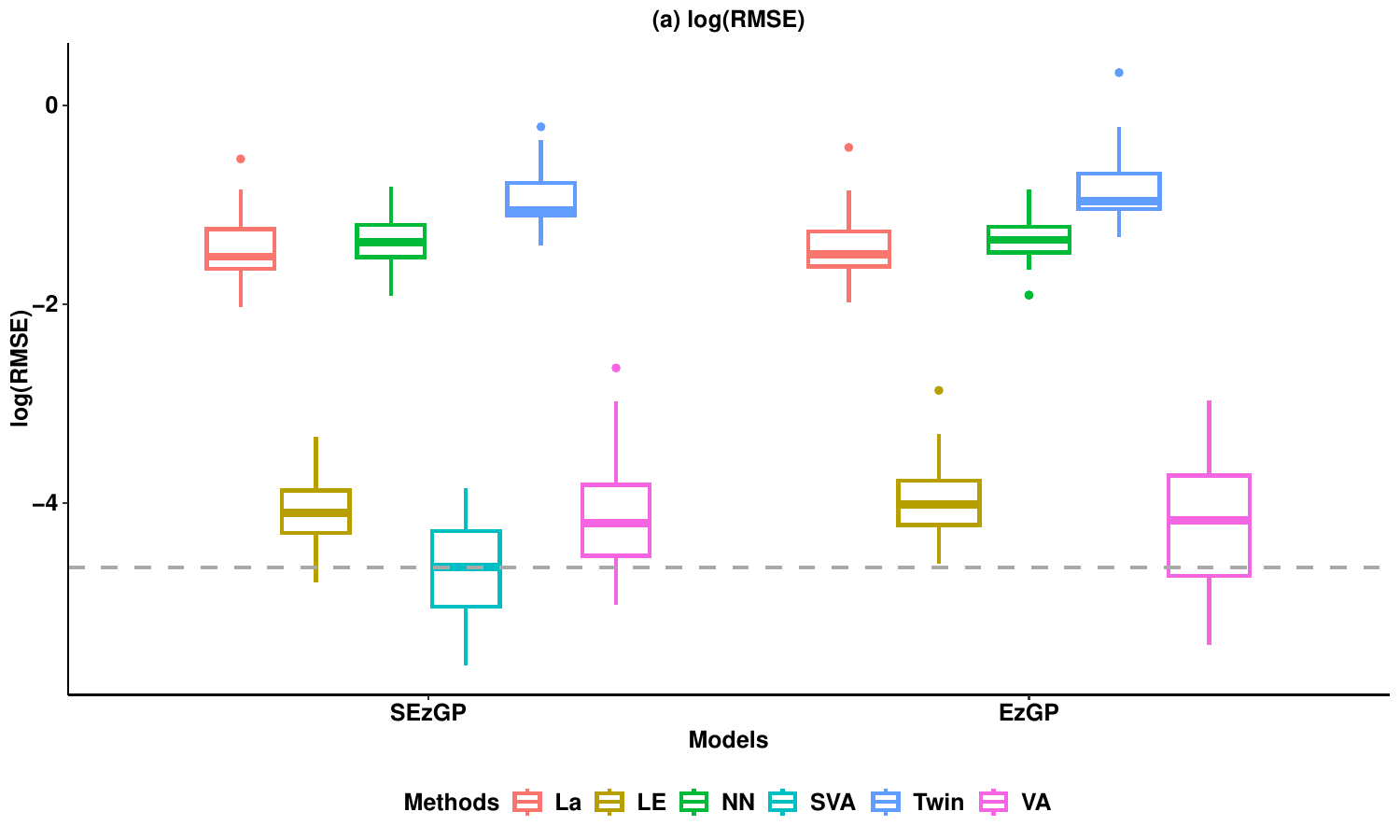}
		\caption{}
		\label{subfig:rmse}
	\end{subfigure}
	
	\vspace{0.5cm} % Adjust spacing between subfigures
	
	\begin{subfigure}[b]{\textwidth}
		\centering
		\includegraphics[width=0.85\textwidth]{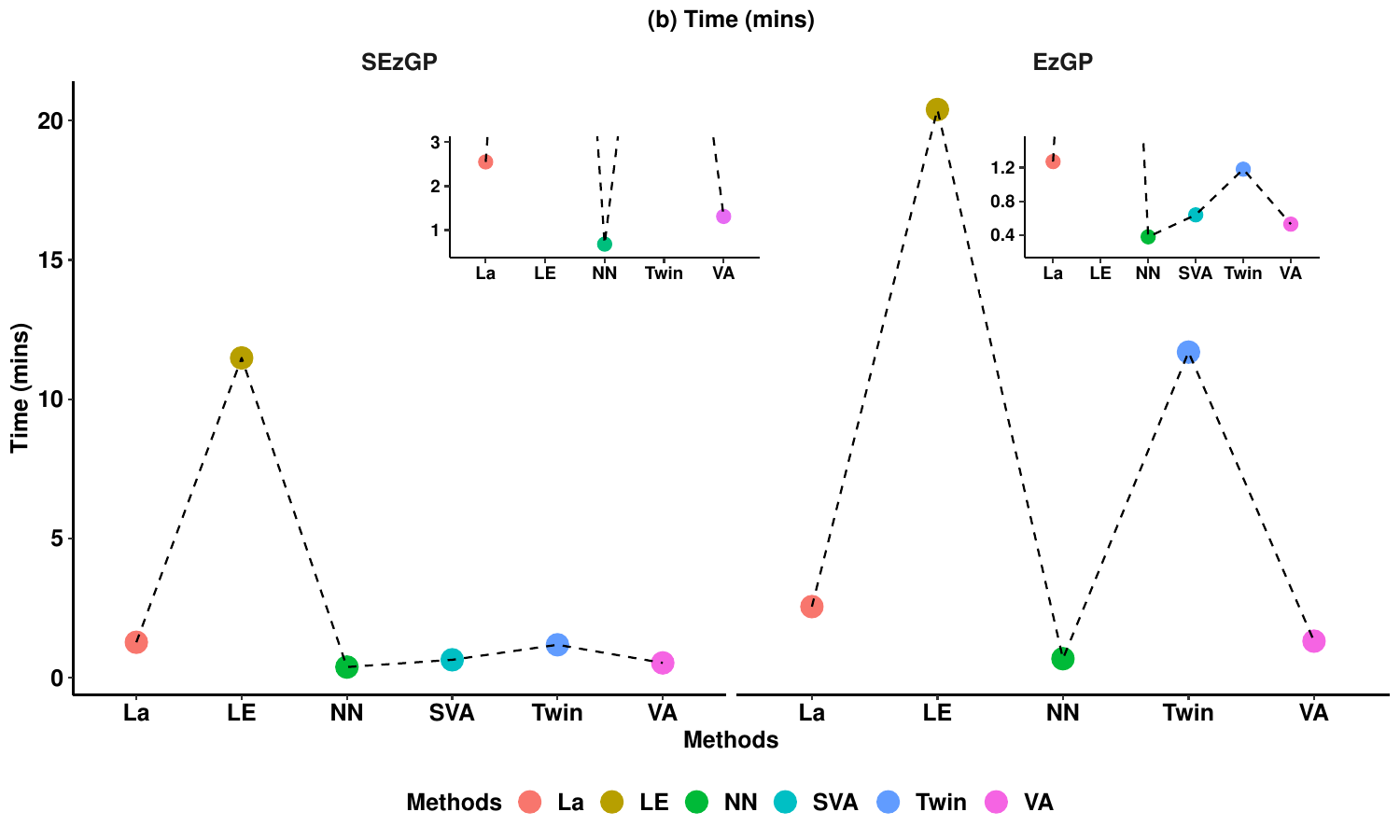}
		\caption{}
		\label{subfig:time}
	\end{subfigure}
	
	\caption{\footnotesize (a) Boxplot of RMSE values; (b) Average computation time in minutes for Twin (105), LE (200), NN, La, VA, and SVA (35) methods, using EzGP and SEzGP across 30 simulations in \nameref{ex4.1}, Scenario 1.}
	\label{box_Ex1_1_E_HE}
\end{figure}

\begin{figure}[!htb]
	\centering
	\begin{subfigure}[b]{\textwidth}
		\centering
		\includegraphics[width=0.85\textwidth]{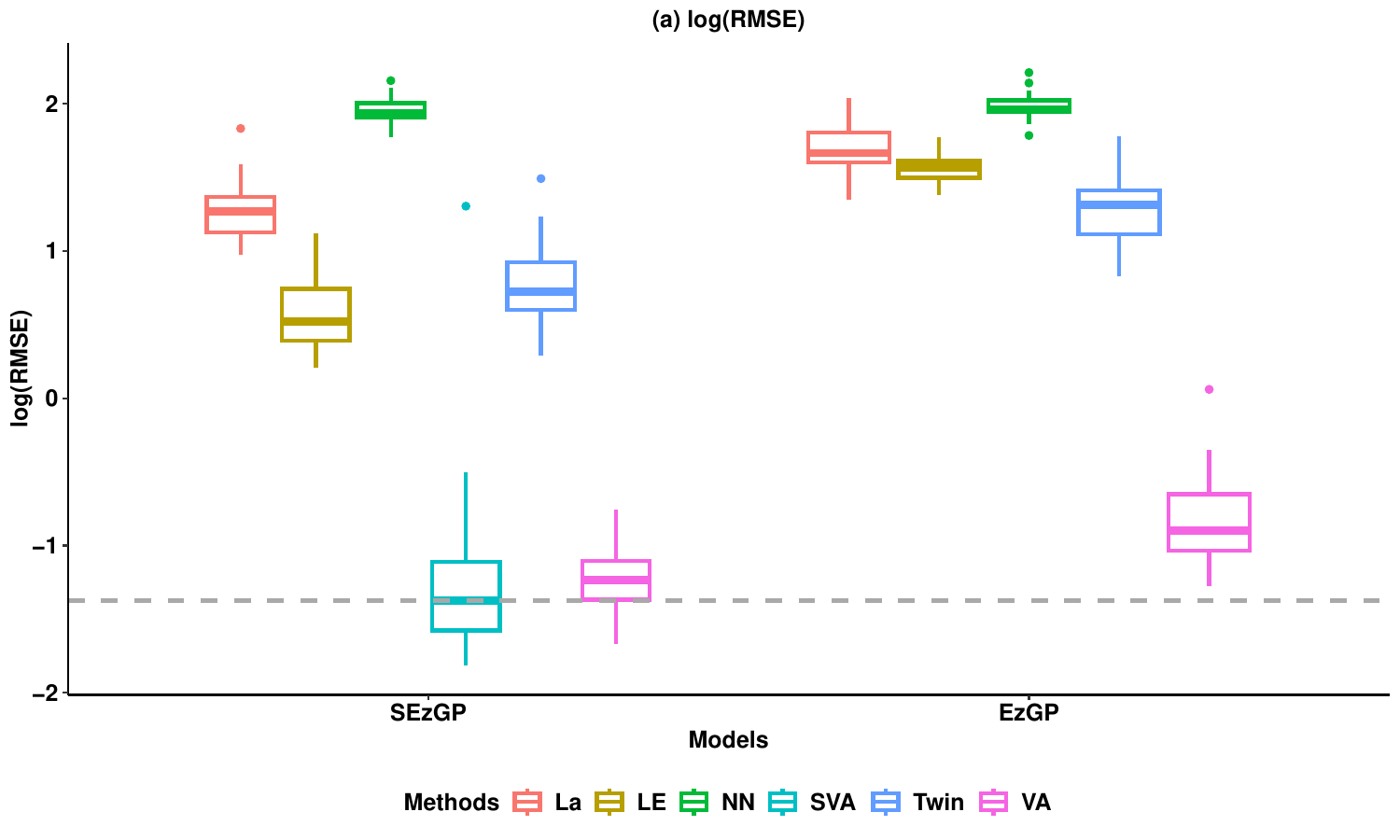}
		\caption{}
		\label{subfig:rmse_ex2}
	\end{subfigure}
	
	\bigskip % Adds vertical spacing between subfigures
	
	\begin{subfigure}[b]{\textwidth}
		\centering
		\includegraphics[width=0.85\textwidth]{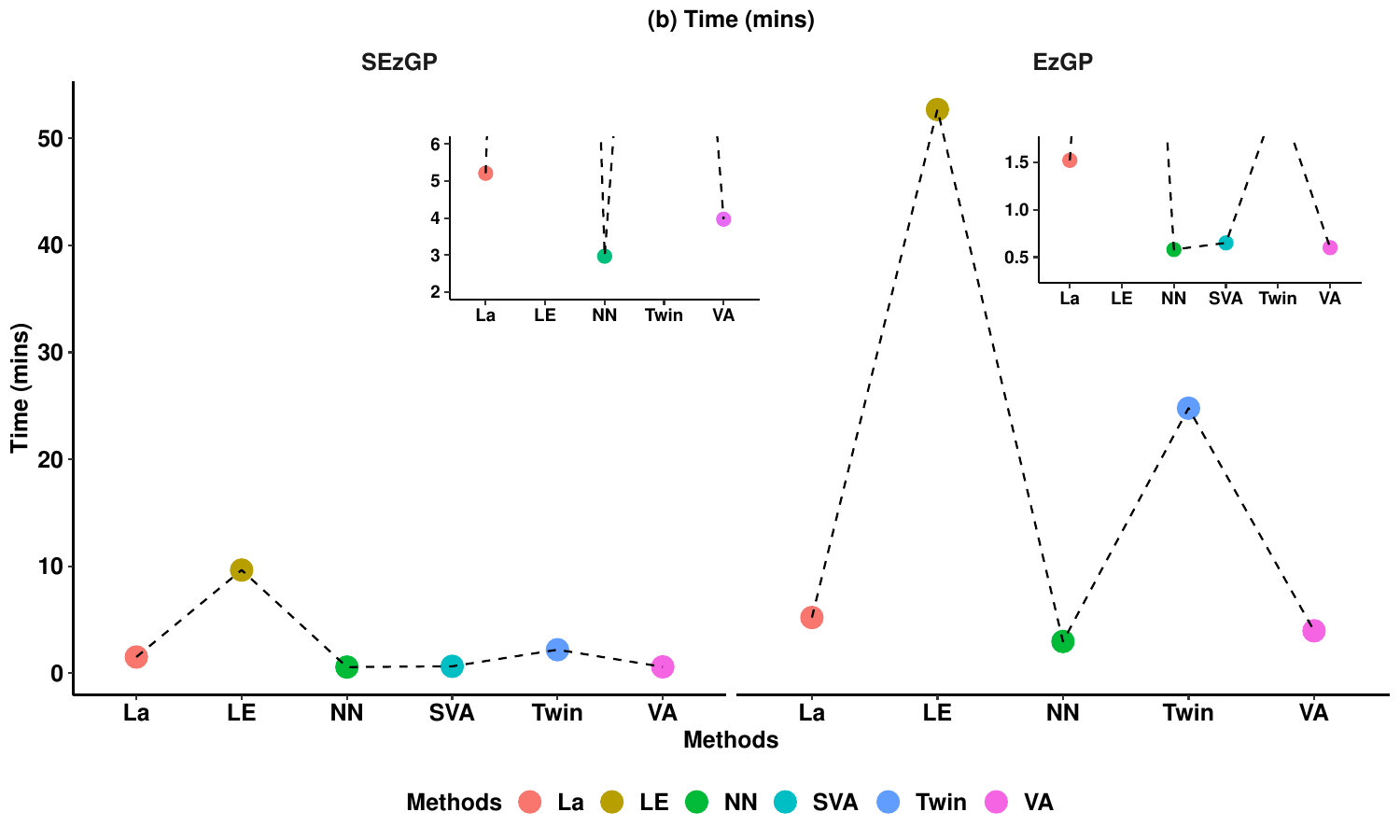}
		\caption{}
		\label{subfig:time_ex2}
	\end{subfigure}
	
	\caption{\footnotesize (a) RMSE boxplot; (b) Average computation time (minutes) for Twin (98), LE (140), NN, La, VA, and SVA (35) methods using SEzGP across 30 simulations in Scenario 2 of \nameref{ex4.1}.}
	\label{box_Ex2_1_E_HE}
\end{figure}

\begin{figure}[!htb]
	\centering
	\begin{subfigure}[b]{\textwidth}
		\centering
		\includegraphics[width=0.85\textwidth]{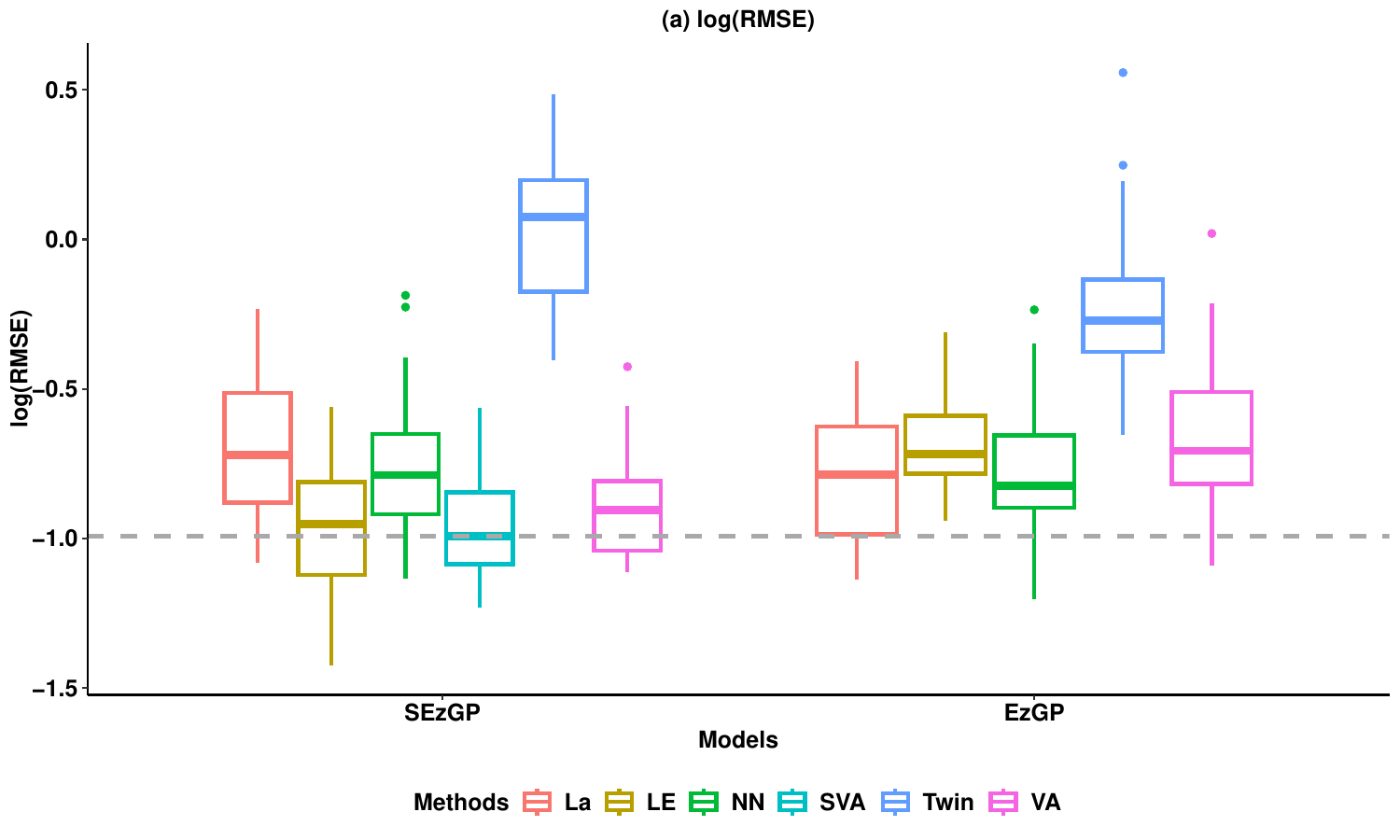}
		\caption{}
		\label{subfig:ex1_2_rmse}
	\end{subfigure}
	
	\vspace{0.5cm}
	
	\begin{subfigure}[b]{\textwidth}
		\centering
		\includegraphics[width=0.85\textwidth]{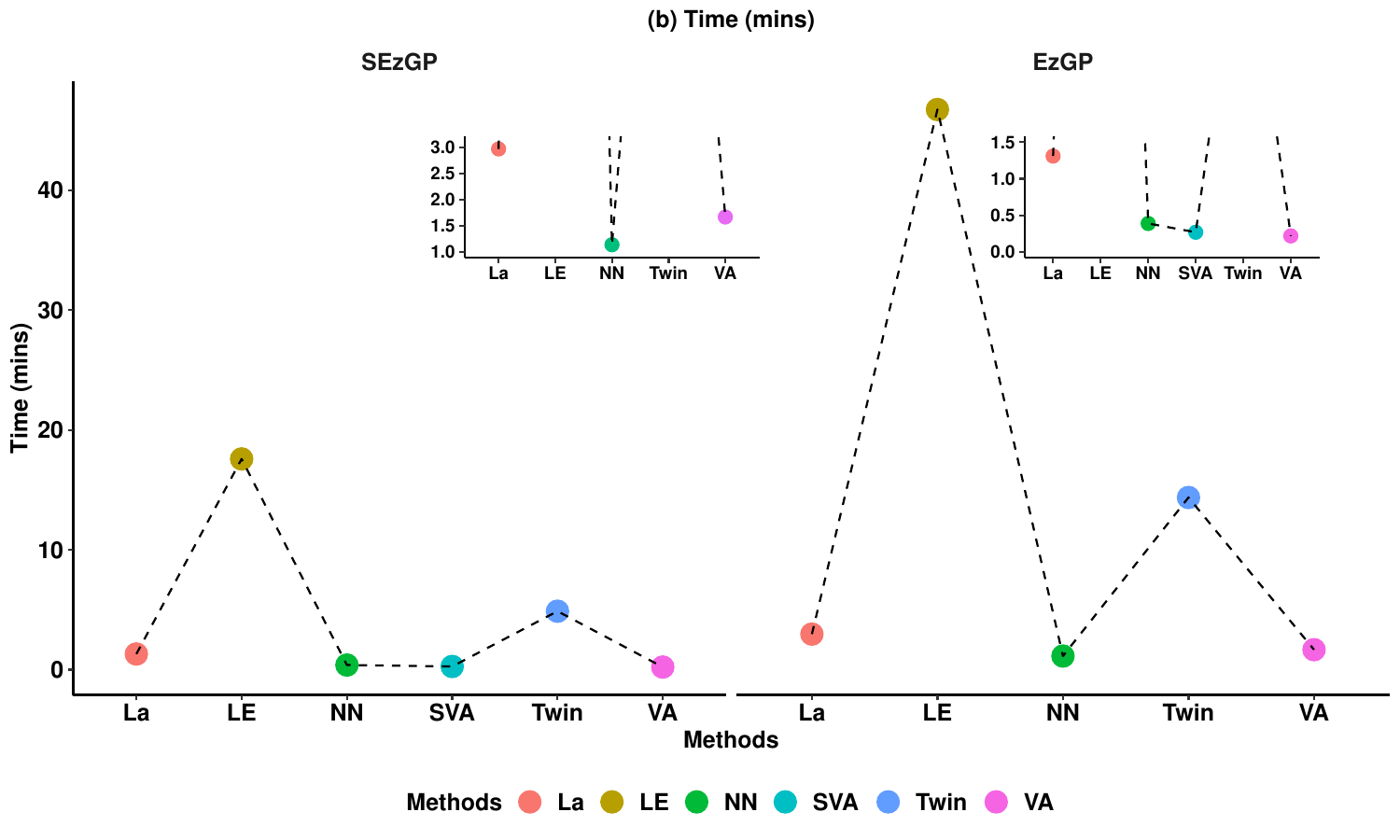}
		\caption{}
		\label{subfig:ex1_2_time}
	\end{subfigure}
	
	\caption{\footnotesize (a) RMSE boxplot; (b) Average computation time (minutes) for Twin (105), LE (200), NN, La, VA, and SVA (35) methods using EzGP and SEzGP across 30 simulations in Scenario 1 of \nameref{ex4.2}.}
	\label{box_Ex1_2_E_HE}
\end{figure}

\begin{figure}[!htb]
	\centering
	\begin{subfigure}[b]{\textwidth} 
		\centering
		\includegraphics[width=0.85\textwidth]{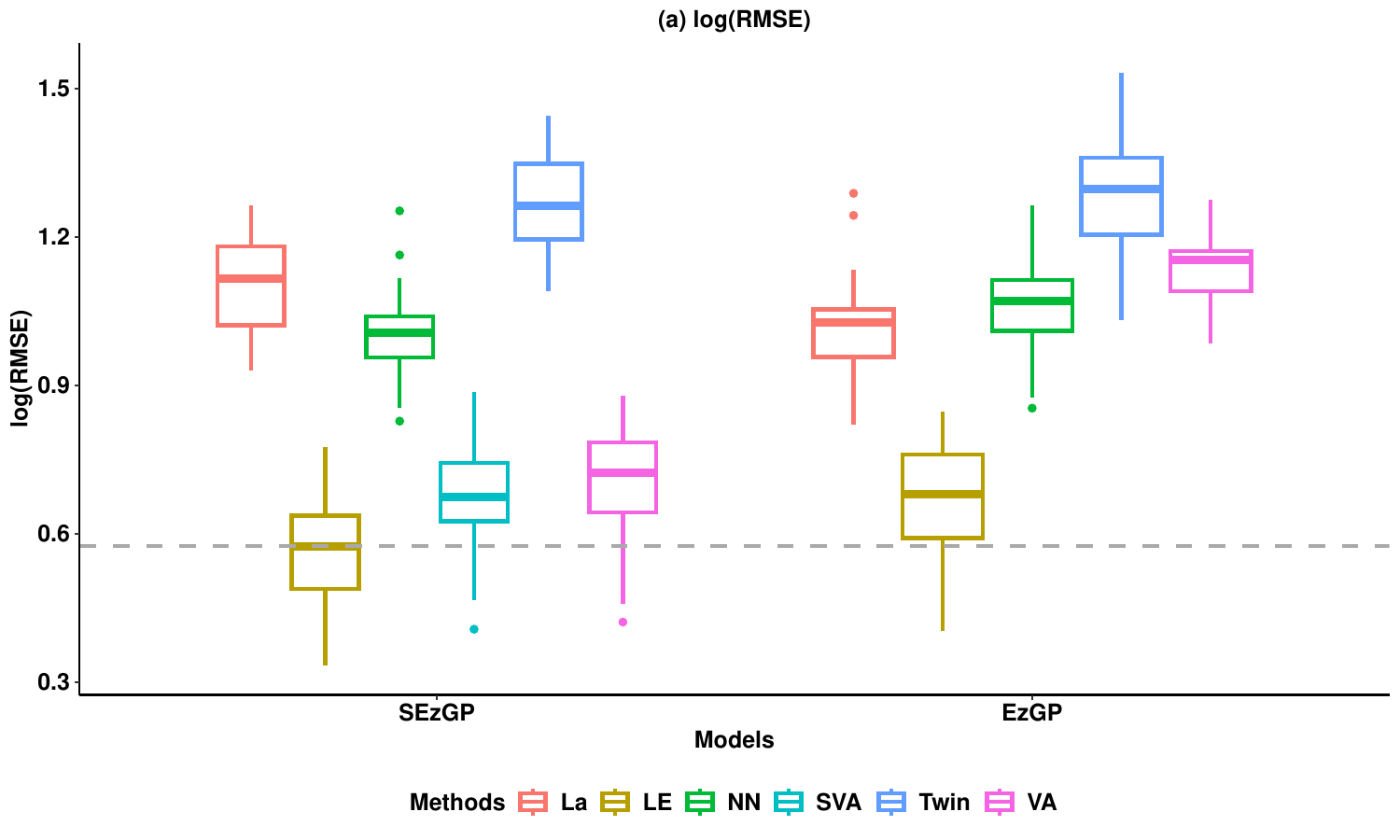}
		\caption{}
		\label{subfig:ex2_2_rmse}
	\end{subfigure}
	
	\vspace{0.5cm}
	
	\begin{subfigure}[b]{\textwidth}
		\centering
		\includegraphics[width=0.85\textwidth]{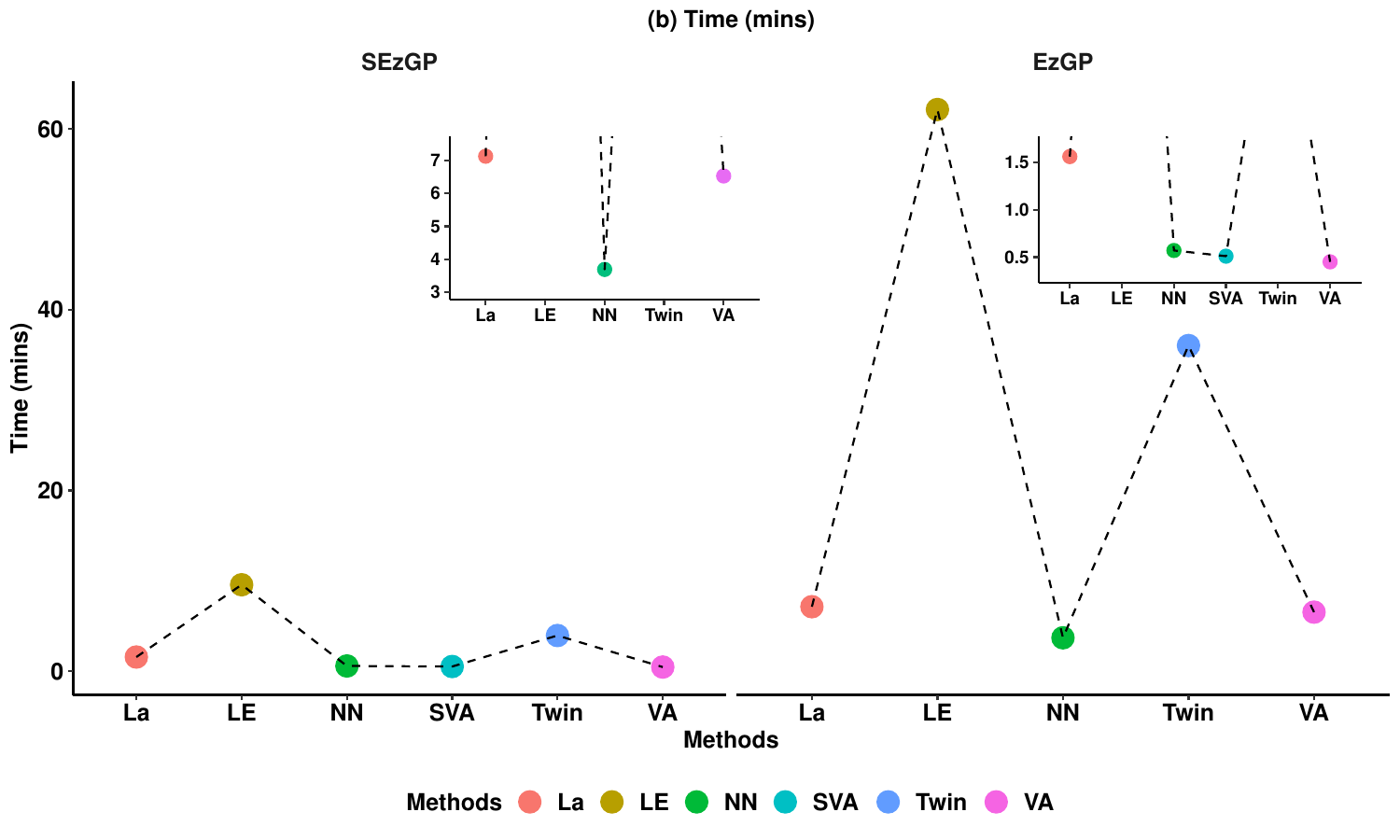}
		\caption{}
		\label{subfig:ex2_2_time}
	\end{subfigure}
	
	\caption{\footnotesize (a) RMSE boxplot; (b) Average computation time (minutes) for Twin (105), LE (140), NN, La, VA, and SVA (35) methods using EzGP and SEzGP across 30 simulations in Scenario 2 of \nameref{ex4.2}.}
	\label{box_Ex2_2_E_HE}
\end{figure}

\begin{figure}[htb!]
	\centering
	\begin{subfigure}[b]{\textwidth}
		\centering
		\includegraphics[width=0.85\textwidth]{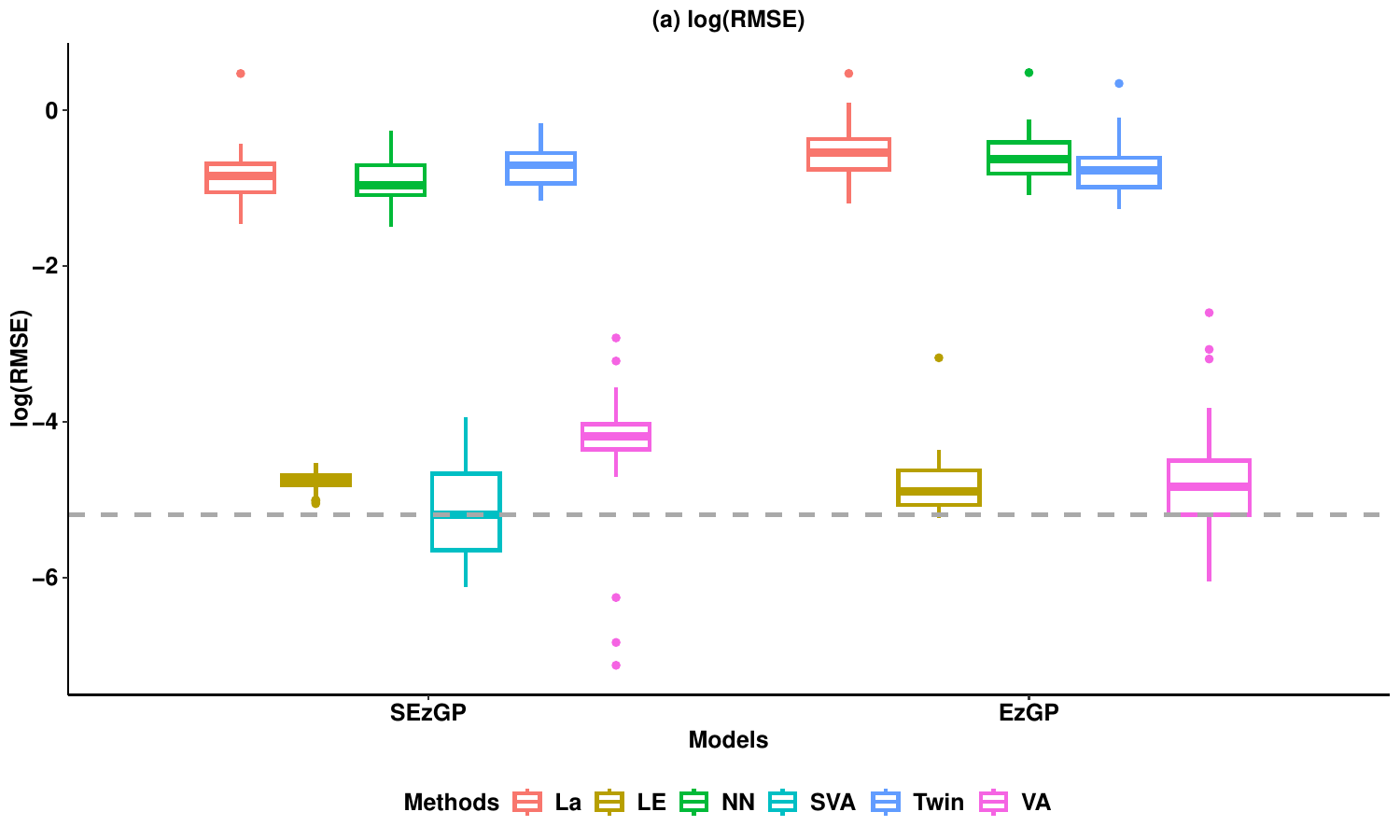}
		\caption{}
		\label{subfig:ex1_3_rmse}
	\end{subfigure}
	
	\vspace{0.5cm}
	
	\begin{subfigure}[b]{\textwidth}
		\centering
		\includegraphics[width=0.85\textwidth]{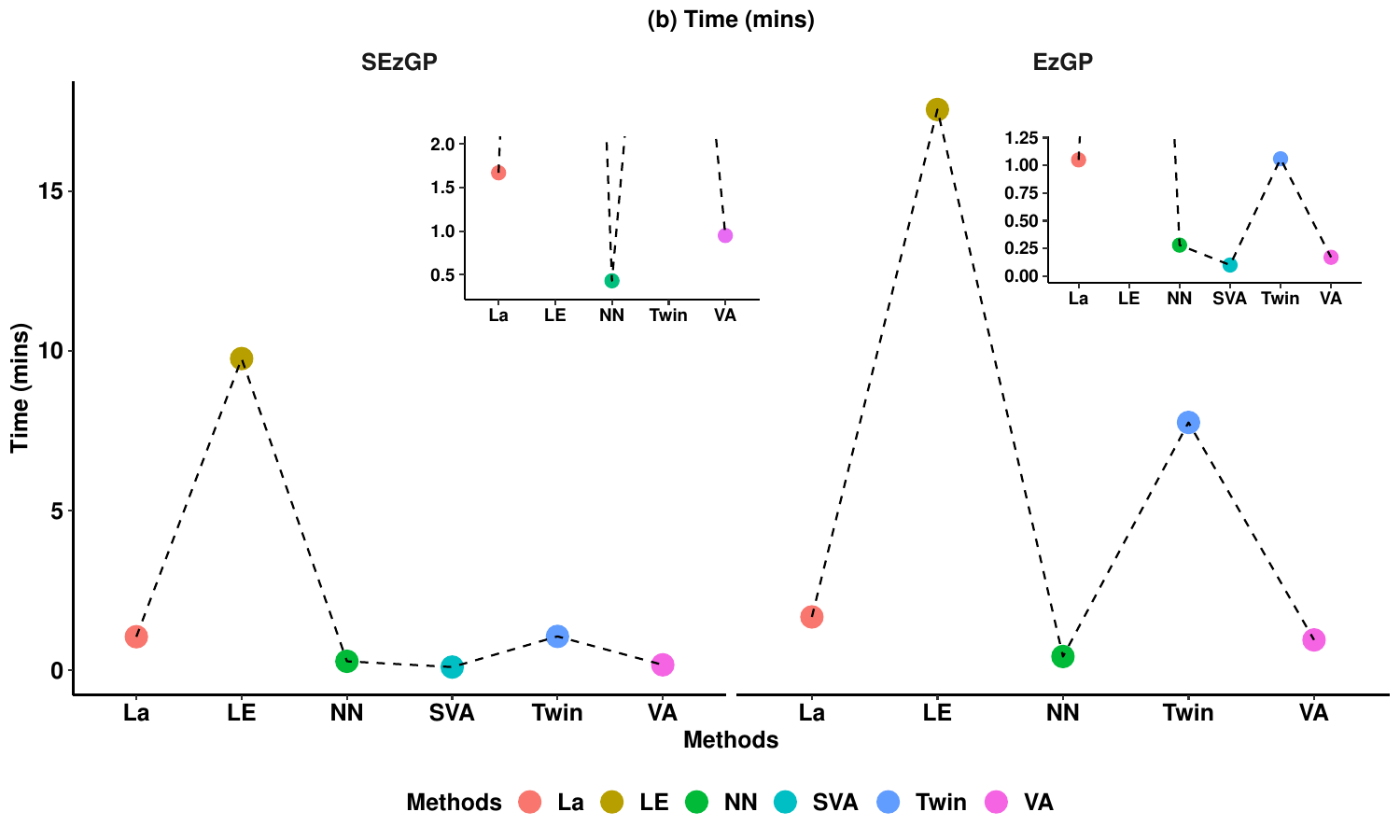}
		\caption{}
		\label{subfig:ex1_3_time}
	\end{subfigure}
	
	\caption{\footnotesize (a) RMSE boxplot; (b) Average computation time (minutes) for Twin (98), LE (200), NN, La, VA, and SVA (35) methods using EzGP and SEzGP across 30 simulations in Scenario 1 of \nameref{ex4.3}.}
	\label{box_Ex1_3_E_HE}
\end{figure}

\begin{figure}[!htb]
	\centering
	\begin{subfigure}[b]{\textwidth}
		\centering
		\includegraphics[width=0.85\textwidth]{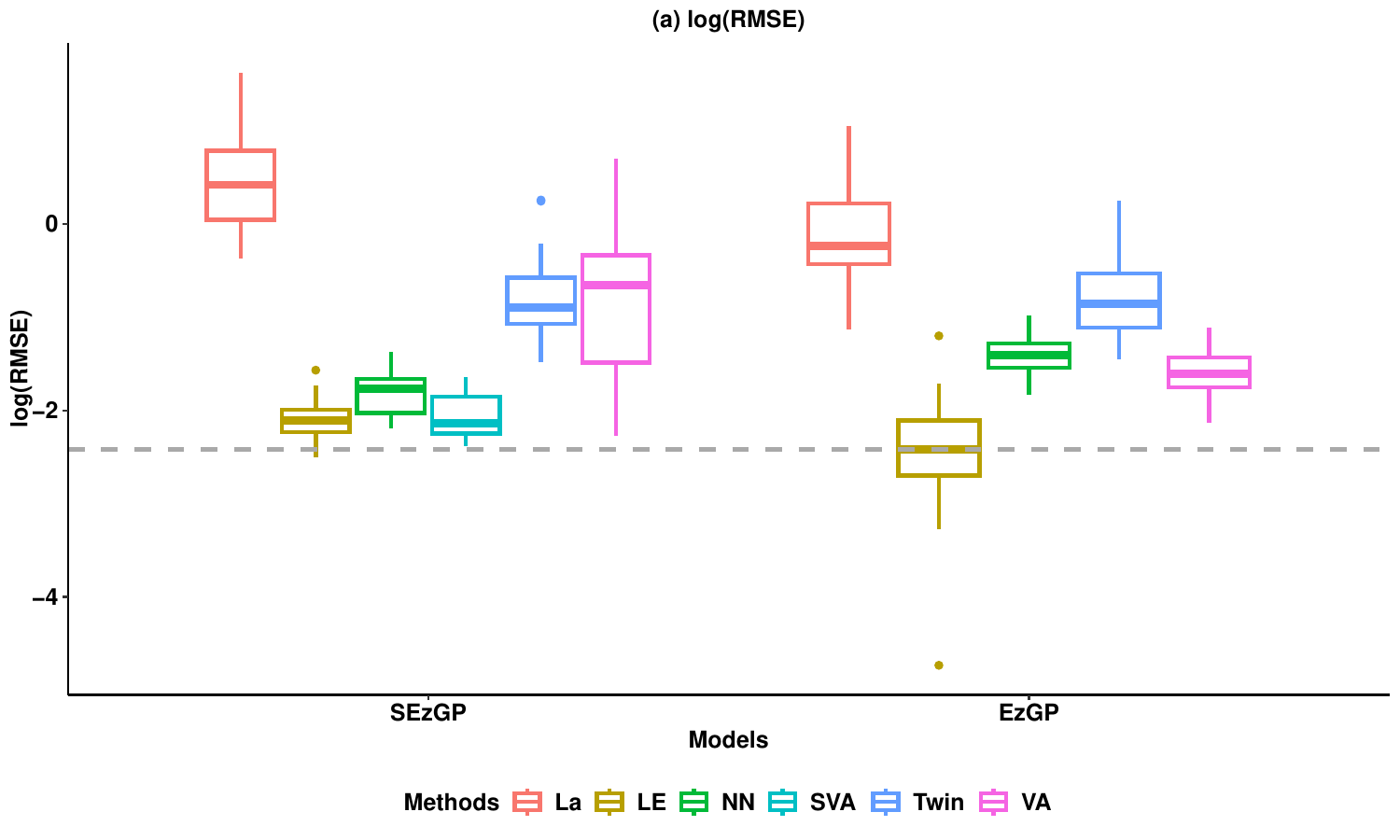}
		\caption{}
		\label{subfig:ex2_3_rmse}
	\end{subfigure}
	
	\vspace{0.5cm}
	
	\begin{subfigure}[b]{\textwidth}
		\centering
		\includegraphics[width=0.85\textwidth]{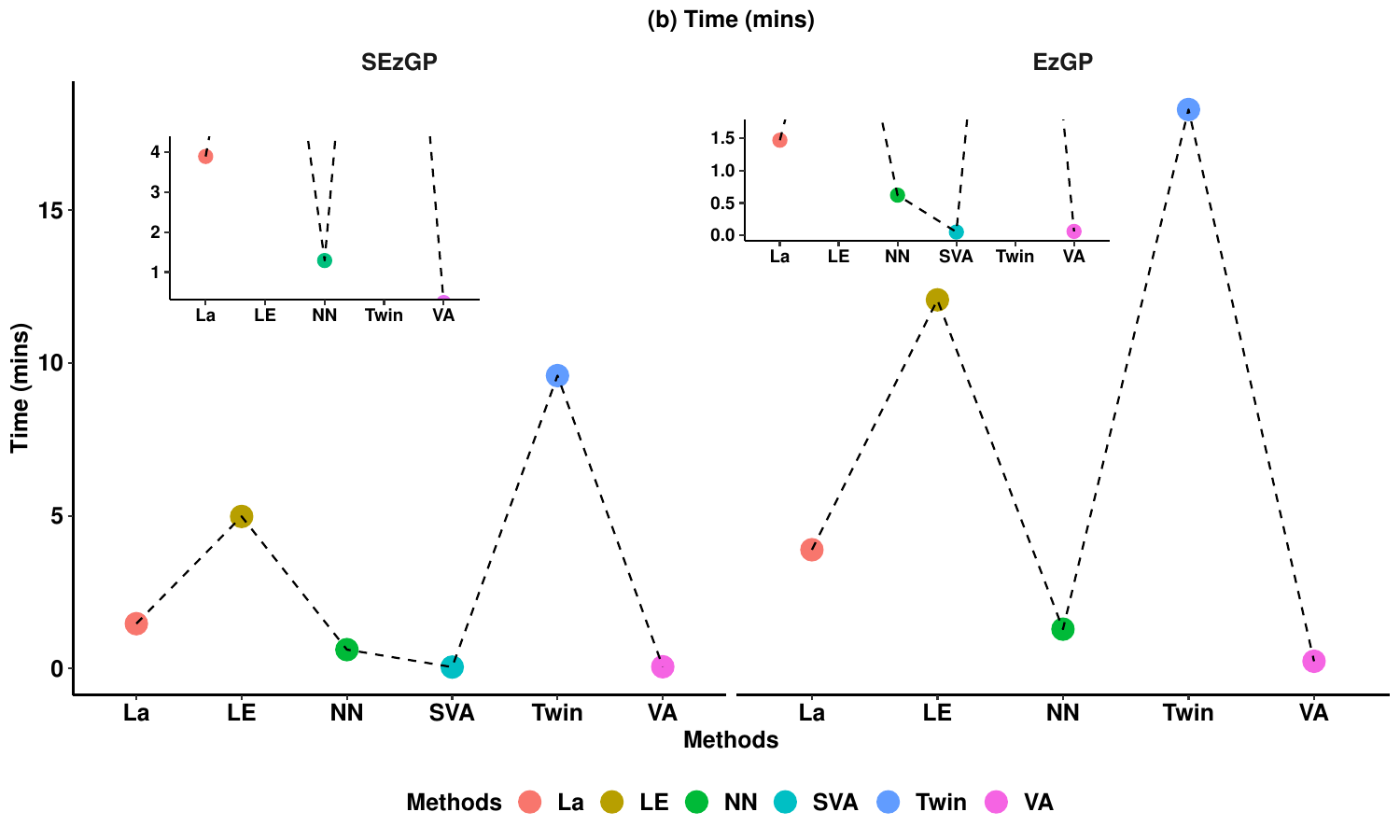}
		\caption{}
		\label{subfig:ex2_3_time}
	\end{subfigure}
	
	\caption{\footnotesize (a) RMSE boxplot; (b) Average computation time (minutes) for Twin (96), LE (65), NN, La, VA, and SVA (35) methods using EzGP and SEzGP across 30 simulations in Scenario 2 of \nameref{ex4.4}.}
	\label{box_Ex2_3_E_HE}
\end{figure}

The SEzGP covariance function not only yields more accurate predictions - as evidenced by lower RMSE values - but also significantly reduces the computational time required for model fitting and prediction. Figures~\ref{box_Ex1_1_E_HE} through~\ref{box_Ex2_3_E_HE} show the logarithm of the RMSE and the average computational time (in minutes) for both covariance functions.

These results demonstrate that SEzGP consistently achieves a better trade-off between predictive accuracy and computational efficiency, making it a more suitable choice for large-scale computer experiments with mixed-inputs.

%\redtext{The figures need to change to have consistent notations and format}

\bibliographystyle{chicago}
\bibliography{reference} % Add your references here

\end{document}